%% file: paper1.tex
\documentclass[aip,reprint]{revtex4-1}

\usepackage{amssymb}
\usepackage{xcolor}
\usepackage{graphicx}
\usepackage{bm}
\usepackage{subcaption}
\usepackage{afterpage}

\newcommand\omegape{\omega_\text{pe}}
\newcommand\omegac{\omega_\text{c}}
\newcommand\lambdad{\lambda_\text{D}}

\input{preamble.tex}
\usepackage{cleveref}

\begin{document}


\title{High-Resolution Particle-In-Cell Simulations of Two-Dimensional Bernstein-Greene-Kruskal Modes}



\author{J.
    McClung} \email[]{james.mcclung@unh.edu}
      \affiliation{Department of Physics, University of
          New Hampshire}

      \author{M.
    T.
    Franciscovich} \affiliation{Geophysical
          Institute, University of Alaska Fairbanks,
          Fairbanks, Alaska 99775, USA}

      \author{K.
    Germaschewski} \affiliation{Department of
          Physics, University of New Hampshire}

      \author{C.
    S.
    Ng}
\email[]{cng2@alaska.edu}
\affiliation{Geophysical Institute, University of Alaska Fairbanks, Fairbanks, Alaska 99775, USA}


\date{\today}

\begin{abstract}
    We present two dimensional (2D) particle-in-cell
      (PIC) simulations of 2D Bernstein-Greene-Kruskal
      (BGK) modes, which are exact nonlinear
      steady-state solutions of the Vlasov-Poisson
      equations, on a 2D plane perpendicular to a
      background magnetic field, with a cylindrically
      symmetric electric potential localized on the
      plane.
    PIC simulations are initialized using analytic
      electron distributions and electric potentials
      from the theory.
    We confirm the validity of such solutions using
      high-resolutions up to a $2048^2$ grid.
    We show that the solutions are dynamically stable
      for a stronger background magnetic field, while
      keeping other parameters of the model fixed, but
      become unstable when the field strength is weaker
      than a certain value.
    When a mode becomes unstable, we observe that the
      instability begins with the excitation of
      azimuthal electrostatic waves that ends with a
      spiral pattern.
\end{abstract}



  \maketitle

  \section{INTRODUCTION
    \label{sec:intro}}

  One
  remarkable property of high-temperature
  collisionless plasmas is the possibility of
  forming small-scale kinetic structures known as
  Bernstein-Greene-Kruskal (BGK) modes,
  \cite{PhysRev.108.546} which can be theoretically
  described as exact nonlinear steady-state
  solutions that satisfy the Vlasov and Poisson
  equations self-consistently.
The original BGK mode solutions are in one
  dimension (1D), with the solution being invariant
  along one Cartesian coordinate.
Analytic solutions of BGK modes were later
  generalized to 3D for an unmagnetized plasma,
  \cite{PhysRevLett.95.245004} with a spherically
  symmetric electric potential localized in all
  spatial directions; as well as 2D, for a
  magnetized plasma in a uniform background
  magnetic field, \cite{doi:10.1063/1.2186187,
      doi:10.1063/1.5126705} with a cylindrically
  symmetric potential localized on a 2D plane
  perpendicular the background field.
Much more research has been done on the subject
  of BGK modes.
Please refer to a recent paper,
  Ref.~\onlinecite{doi:10.1063/1.5126705}, for a
  more complete discussion on physical motivations
  and relevant literature along this line of
  research.

Preliminary low-resolution 2D particle-in-cell
  (PIC) simulations of 2D BGK modes
  \cite{doi:10.1063/1.2186187,
      doi:10.1063/1.5126705} have been reported before,
  \cite{doi:10.1063/1.4723590} showing a
  qualitative trend that the modes are more stable
  in the limit of strong background magnetic field
  strength, and unstable in the opposite limit.
In this paper, we report 2D PIC simulations of
  such 2D BGK modes in much higher resolutions
  (using grids of cells up to $2048^2$), using the
  Plasma Simulation Code (PSC),
  \cite{GERMASCHEWSKI2016305} a fully
  electromagnetic PIC code, running in parallel on
  supercomputers.

The main objectives of this research are to
  confirm the validity of analytic solutions,
  \cite{doi:10.1063/1.2186187,
      doi:10.1063/1.5126705} quantify how the stability
  of such modes depends on key parameters such as
  the background magnetic field strength, and to
  observe the dynamical evolutions of the modes
  after they become unstable.
In \cref{sec:theory}, we briefly summarize the
  theory of the 2D BGK modes.
\cite{doi:10.1063/1.2186187, doi:10.1063/1.5126705}
In \cref{sec:setup}, we describe the numerical
  setup of the PIC simulations.
Main results from these simulations are presented
  in \cref{sec:results}.
Based on these results, we discuss and conclude
  in \cref{sec:conclusion} on how the above
  objectives are met.
In short, our results will show that the analytic
  solutions of the 2D BGK modes are indeed valid
  for the Vlasov-Poisson equations, and that such
  solutions are stable in 2D for our choice of
  parameters with the normalized strength of the
  background magnetic field $B_0 \gtrsim 2$
  (defined in the next section) for our choice of
  other parameters, while azimuthal electrostatic
  waves are excited when a mode becomes unstable.

\section{BGK MODE SOLUTIONS \label{sec:theory}}
\subsection{Theory}

The Vlasov equation is the continuity equation for a distribution function $f$ over position $(x_1, x_2, x_3)$ and velocity $(v_1, v_2, v_3)$ space,
\begin{equation}
    \pderiv ft + \sum_{i=1}^3\pderiv f{x_i}\deriv{x_i}t + \sum_{i=1}^3\pderiv f{v_i}\deriv{v_i}t = 0.
\end{equation}

We consider the case with a uniform ion density
  $n_i$ and a uniform background magnetic field
  $\vec B_0 = B_0\unit z$.
In cylindrical coordinates,
  $(x_1,x_2,x_3;v_1,v_2,v_3) = (\rho,\phi,z;
      v_\rho, v_\phi, v_z)$.

We normalize according to
\begin{subequations}
    \label{eq:normalized 1}
    \begin{align}
        e\ideq q_i = -q_e                                  & \to 1                     \\
        m_e                                                & \to 1                     \\
        n_i=n_e\rvert_{\rho=\infty}                        & \to 1                     \\
        \epsilon_0                                         & \to 1                     \\
        \omegape \ideq \sqrt{\frac{e^2n_e}{\epsilon_0m_e}} & \to 1 \label{eq:normt}    \\
        v_e\ideq \sqrt{\frac{k_BT_e}{m_e}}                 & \to 1 \label{eq:normv}  .
    \end{align}
\end{subequations}

The normalizations \eqref{eq:normt} and
  \eqref{eq:normv} imply that the Debye length
  $\lambdad\ideq v_e/\omegape = 1$.
The electric potential $\psi$ has units
  $m_ev_e^2/e$, and the magnetic field strength
  $B_0$ has units $em_e/n_e\omegape$.
Note that $B_0=\omegac/\omegape$, where
  $\omegac\ideq eB/m_e$ is the electron
  gyrofrequency.
Thus, the electron gyroperiod $\tau_c\ideq
      2\pi/\omegac$ is $2\pi/B_0$.

Imposing spatial symmetries $\pderiv{}\phi=\pderiv{}z=0$ and the steady-state condition $\pderiv{}t=0$ on the Vlasov-Poisson equations obtains
\begin{align}
    \begin{split}
        \pderiv{f}\rho v_\rho + \pderiv{f}{v_\rho}\left(\pderiv\psi\rho - B_0v_\phi + \frac{v_\phi^2}\rho\right) + &                       \\
        \pderiv{f}{v_\phi}\left(B_0v_\rho - \frac{v_\rho v_\phi}\rho\right)                                        & = 0,\label{eq:vlasov}
    \end{split} \\
    \frac1\rho\pderiv{}\rho \left(\rho\pderiv\psi\rho\right) = \int\dd^3v f - 1.\label{eq:poisson}
\end{align}

When $f$ depends only on energy
  $w=\frac12v^2-\psi$, it cannot satisfy
  \eqref{eq:poisson} and remain localized.
\cite{doi:10.1063/1.5126705}
Including a dependence on the canonical angular
  momentum $l=\rho v_\phi - \frac12 B_0\rho^2$
  makes a localized solution possible.
One possible form is
\begin{align}
    f(w,l) = (2\pi)^{-3/2}e^{-w}\left(1-he^{-kl^2}\right) \label{eq:exact dist}
\end{align}
where $h\in(-\infty,1)$ and $k\in(0,\infty)$ are constant parameters. The potential $\psi=\psi(\rho)$ must be solved numerically using Eq.~(\ref{eq:poisson}), with $f$ given by Eq.~(\ref{eq:exact dist}),
\begin{align}
    \frac1\rho\pderiv{}\rho \left(\rho\pderiv\psi\rho\right) = e^\psi \left[1 - \frac{h}{\gamma}\exp\left(-\frac{kB_0^2\rho^4}{4\gamma^2}\right)\right] - 1 \label{eq:final poisson}
\end{align}
where
\begin{align}
    \gamma \ideq \gamma(\rho) & \defeq \sqrt{1 + 2k\rho^2}
\end{align}

The resulting configuration can be described as
  follows.
Starting with a uniform distribution, electrons
  near the origin have been scooped out and
  deposited in a ring around the origin.
This leads to an outward $E_\rho$.
Additionally, the electrons in the ring have been
  given a slight clockwise flow.
The configuration can therefore be described as
  an electron vortex circulating around an electron
  density hole.
Figures \ref{fig:seqs_max} and
  \ref{fig:seqs_exact} illustrate electron
  distributions for different values of $B_0$ at
  time $t=0$ and later.

\subsection{Approximations} \label{sec:approximations}
These solutions assume a static magnetic field
  $\vec B=B_0\hat x$.
Induced magnetic fields due to plasma current
  were negligible even for $B_0=0.1$, the lowest
  value of $B_0$ considered in this paper.
This is due to our choice of the parameter
  $\beta_e = v_e/c = 0.001$.

Although the plasma in these simulations is very
  cold, with $k_BT=0.511$ eV, and is magnetically
  dominated with a plasma beta of $2\times10^{-6}$
  when $B_0=1$, the solution given by
  Eqs.~\eqref{eq:exact dist} and \eqref{eq:final
      poisson} does not depend on $\beta_e$ explicitly.
Therefore, they are valid if a larger $\beta_e$
  value is used, as long as the physics remains in
  the non-relativistic regime.
The small $\beta_e$ value used in the
  relativistic PSC simulations is to make sure the
  outputs can be compared with the non-relativistic
  theory.
For larger $\beta_e$, Amp\`{e}re's Law would need
  to be solved alongside Eq.~\eqref{eq:final
      poisson} to obtain a self-consistent magnetic
  field.
\cite{doi:10.1063/1.5126705}
For the relativistic case with $\beta_e
      \rightarrow 1$, the relativistic Vlasov equation
  would need to be employed instead.
These generalizations are left for future
  studies, and will not be considered in this
  paper.

An ostensibly less justifiable approximation was
  that ions formed a uniform static background.
It is possible to ease this restriction and find
  theoretical solutions with realistic proton
  behavior,\cite{tang2020two} and several runs of
  this nature were performed.
However, such runs were found to be qualitatively
  similar to motionless-ion runs, and we will not
  be exploring them further in this paper.

\section{NUMERICAL SETUP \label{sec:setup}}

\subsection{Normalization \label{sec:psc normalization}}
PSC and figures shown below are normalized according to Eq.~\eqref{eq:normalized 1} with the change
\begin{align}
    c & \to 1.
    \label{eq:normalized 2}
\end{align}
Thus, time still has units of $\omegape^{-1}$,
  but distance now has units of the electron skin
  depth, $c\omegape^{-1}$.
This multiplies the length scale compared to
  Eq.~\eqref{eq:normalized 1} by a factor of
  $\beta_e =v_e/c=0.001$.

\subsection{Parameters}

All runs used constants $h=0.9$, $k=0.4$ (see
  Eq.~\eqref{eq:exact dist}) and electron thermal
  velocity $v_e=c/1000$.
Equation \eqref{eq:final poisson} was solved
  numerically in an external program to obtain
  $\psi(\rho)$.
Ions had unit charge and masses of $10^9$ to
  mimic a static, uniform background of normalized
  density $n_i=1$.

The physical domain for each run was square
  centered at the origin.
Boundaries were periodic.
The side length of the domain was automatically
  fine-tuned for each value of $B_0$ to be large
  enough to avoid strong edge effects, but no
  larger, since resolving the central hole became
  difficult for $B_0\lesssim1$.
The lengths are given in Table \ref{tab:sizes}.

\begin{table}[htb]
    \caption{\label{tab:sizes}
        Side lengths of the square domain used in
          simulations for different values of $B_0$.
    }
    \begin{tabular}{lr}
        $B_0$ & length   \\
        \hline
        0.1   & 0.180062 \\
        0.25  & 0.072156 \\
        0.5   & 0.036307 \\
        1     & 0.020960 \\
        2     & 0.017530 \\
        4     & 0.015464 \\
        10    & 0.013264
    \end{tabular}
\end{table}

\subsection{Field and Density Initialization\label{sec:field init}}

Since $\psi(\rho)$ was solved externally, its
  discretized form had to be transferred to PSC by
  file.
Support for input-by-file was implemented in PSC
  for this purpose.
PSC then interpolates $\psi(\rho)$ onto its grid
  and computes the first-order discrete gradient to
  find the electric field.

Electron number density $n_e$ is computed as
  $n_i=1$ minus the first-order discrete divergence
  of $\vec E$.
For both electrons and ions, a number density of
  1 corresponds to 100 macroparticles per cell
  (\code{nicell}=100) unless otherwise specified.
This means that higher-resolution runs generally
  had more macroparticles.

The magnetic field was simply initialized to
  $\vec{B}(t=0)=B_0\hat{x}$ at every grid point.

\subsection{Velocity Initialization\label{sec:vel init}}

\begin{figure}
    \centering
    \includegraphics[width=\columnwidth]{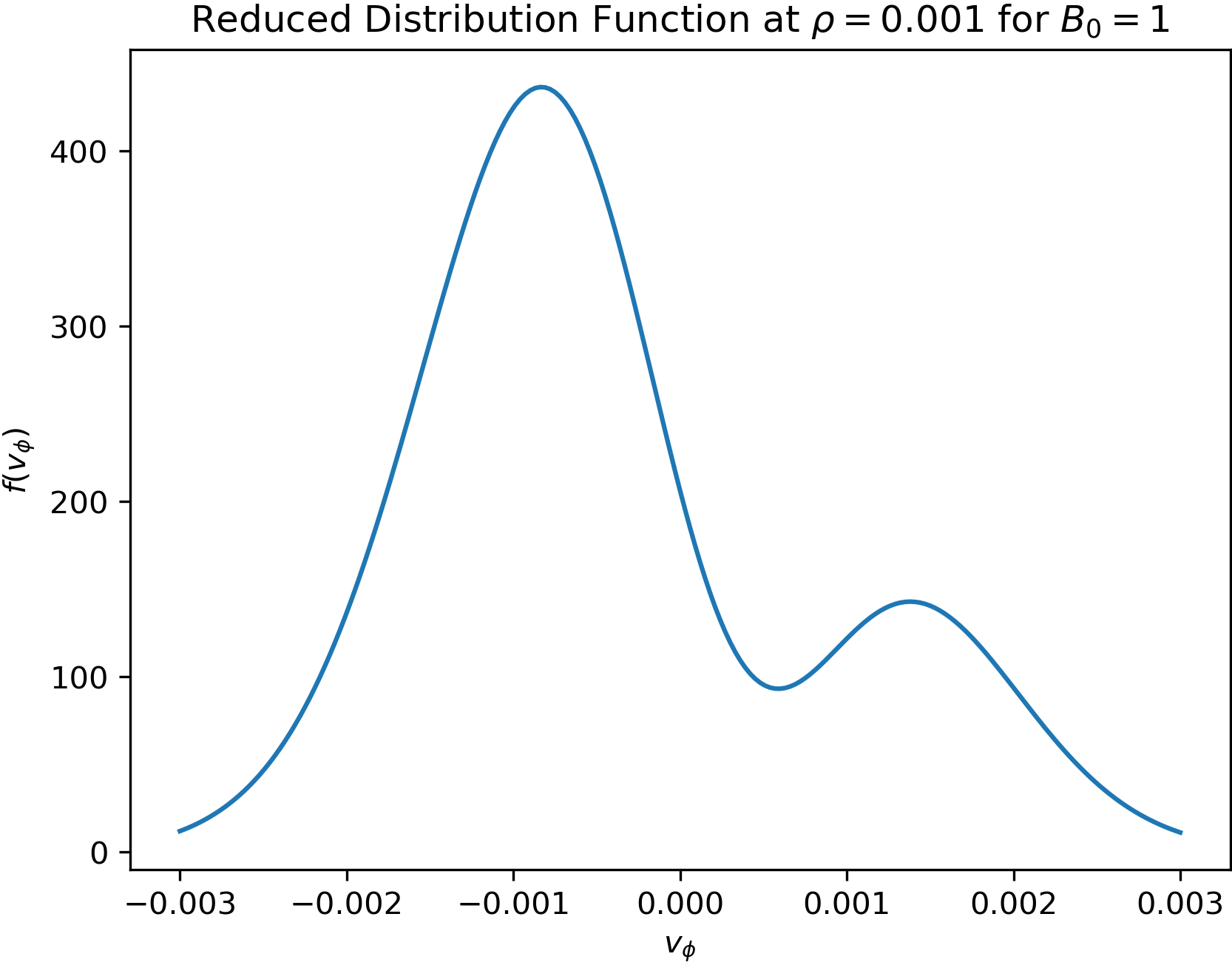}
    \caption{Reduced electron distribution function $f(v_\phi)$ at a point one Debye length from the origin for the $B_0=1$ case.
        All cases feature non-Maxwellian azimuthal
          velocity distributions, but local minima appear
          for $B_0\lesssim2$, with other parameters fixed.
        \label{fig:reduced dist}}
\end{figure}

A major limitation of PSC was that it used the
  Box-Muller method to sample velocities during
  particle initialization.
This method is efficient but produces a normal
  distribution with a given mean and variance.
The radial component of velocity is indeed
  normally distributed according to \eqref{eq:exact
      dist}, but the azimuthal component $v_\phi$ is
  not, as shown in Fig.~\ref{fig:reduced dist}.
Note that the unit of velocity is the speed of
  light $c$ in this figure, and other figures of
  this paper, following the default unit used in
  PSC.

Early runs, called ``moment" runs, simply
  approximated the true distribution as a local
  Maxwellian with the same density, mean velocity,
  and temperature.
We later extended PSC to support arbitrary
  initial distribution functions using an
  inverse-CDF method.
Runs initialized using this method are called
  ``exact" runs.

\section{NUMERICAL RESULTS \label{sec:results}}

Four categories of runs were performed: moment,
  moment reversed, exact, and exact reversed.
``Reversed" cases, intended to serve as non-examples to contrast with the steady-state cases, were initialized exactly the same as non-reversed cases except that $v_\phi$ was negated.
Many runs were performed within each category,
  varying $B_0$ from $0.1$ to $10$ and on grids of
  size $256^2$, $512^2$, and for some cases up to
  $2048^2$.

Note that in the figures of this paper, we use
  the electron skin depth $c/\omega_{pe}$ as the
  unit of spatial length, which is the default unit
  used in PSC, and corresponds to $10^3$ Debye
  lengths for our choice of $v_e$ as discussed in
  \cref{sec:psc normalization}.
From Fig.~\ref{fig:seqs_max} we see that the
  electron density holes have a size on the order
  of $\lambda_D$.
Also following the convention of PSC, the
  background magnetic field is along the
  $x$-direction, so the 2D simulation plane is the
  $y$-$z$ plane.

\subsection{Moment Runs}

\begin{figure*}
    \centering
    \begin{subfigure}[b]{\textwidth}
        \centering
        \includegraphics[width=\textwidth]{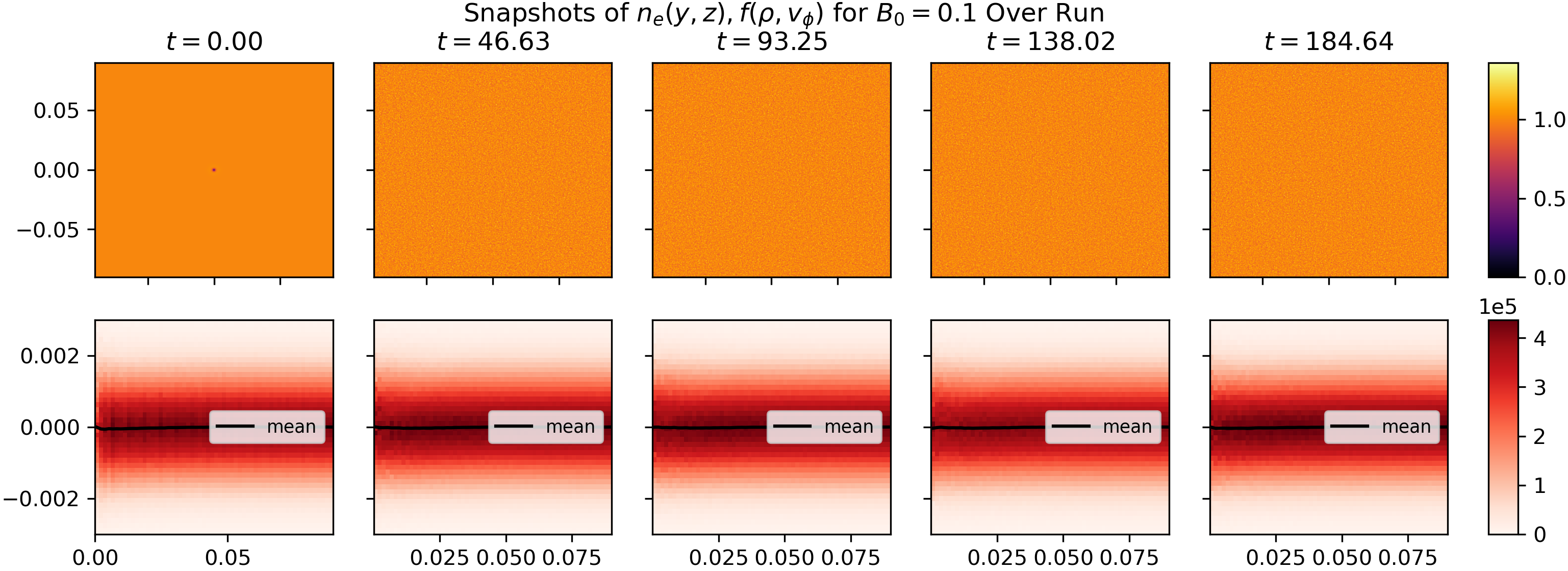}
    \end{subfigure}
    \begin{subfigure}[b]{\textwidth}
        \centering
        \includegraphics[width=\textwidth]{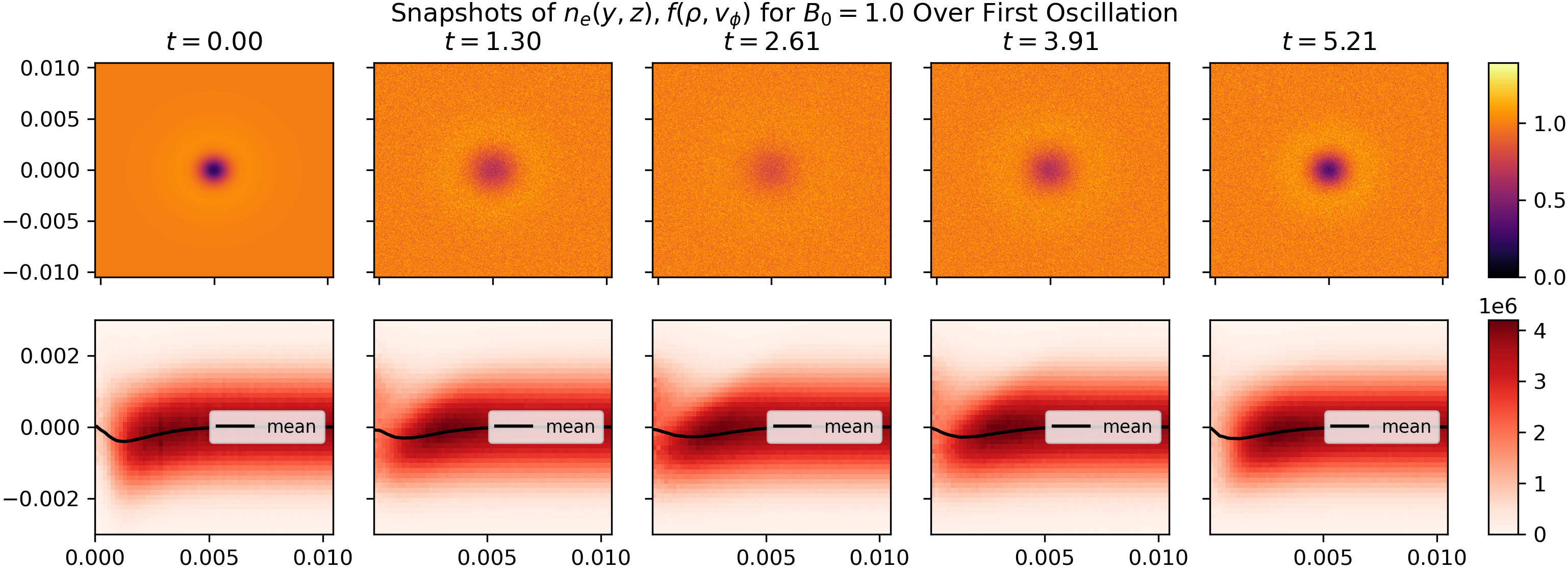}
    \end{subfigure}
    \begin{subfigure}[b]{\textwidth}
        \centering
        \includegraphics[width=\textwidth]{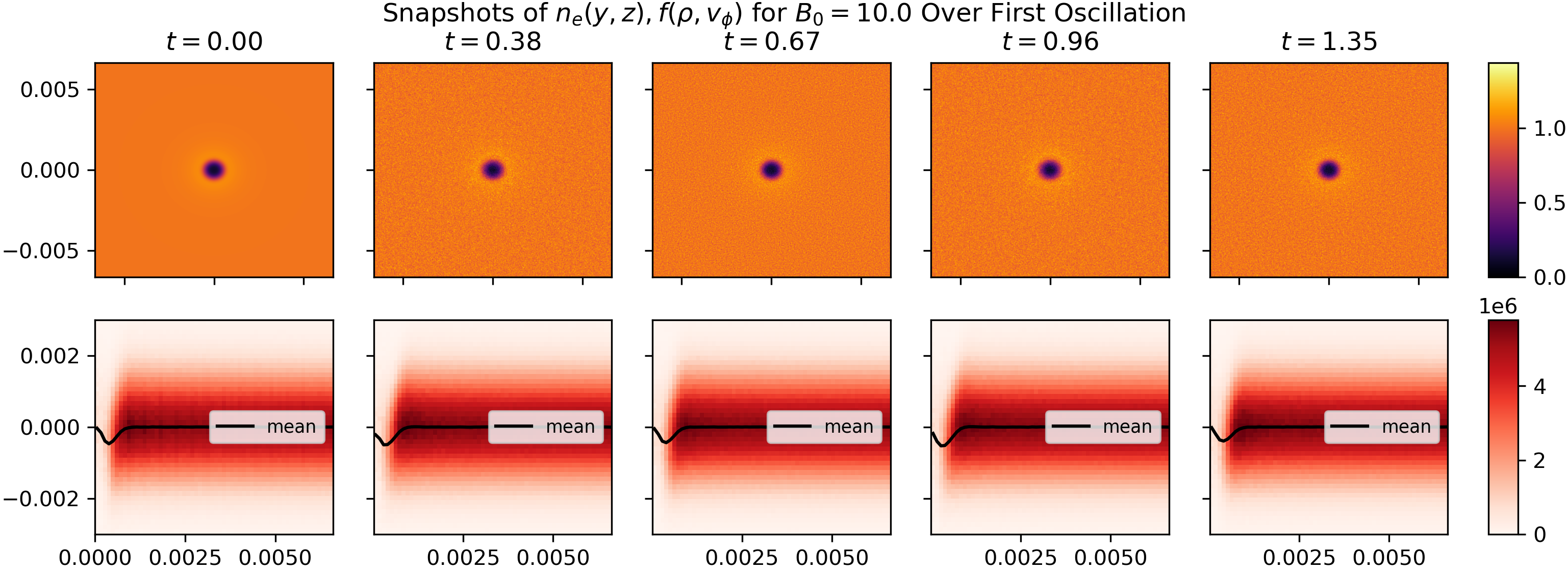}
    \end{subfigure}
    \caption{Sequences of $f(y,z)$ and $f(\rho, v_\phi)$ for moment runs over one oscillation for $B_0\ge1$, or over the whole run for $B_0=0.1$ which did not oscillate.
        Note the pulsation for $B_0=1$ involves a
          collapse of the hole and formation of a ``spike"
          in velocity space, discussed in \cref{sec:spike}.
        \label{fig:seqs_max}}
\end{figure*}

Due to technical limitations of PSC (as discussed
  in \cref{sec:vel init}), early runs used a
  moment-based approximation of the electron
  velocity distribution.
For $B_0\gtrsim4$, runs were more or less
  steady-state, remaining stable for the duration
  of the simulations.
Moment runs with $B_0\lesssim2$, however, were
  not steady-state.
Pulsation was observed for many cases, and are
  more obvious for $0.5\lesssim B_0\lesssim2$.
Runs with $B_0\lesssim0.25$ completely decayed in
  a short period of time.
Fig.~\ref{fig:seqs_max} shows snapshots of low-,
  mid-, and high-$B_0$ moment runs demonstrating
  these phenomena.

In contrast to normal moment runs with high
  $B_0$, moment reversed runs with high $B_0$
  pulsated strongly.
No moment reversed run was steady-state, as
  expected.
For each $B_0$, the reversed run breathed more
  strongly than the non-reversed run (except for
  $B_0=0.1$, which decayed quickly in both cases).

\subsubsection{Nearly Stable: $B_0=4$, Moment\label{sec:stable-moment}}

\begin{figure}
    \centering
    \includegraphics[width=\columnwidth]{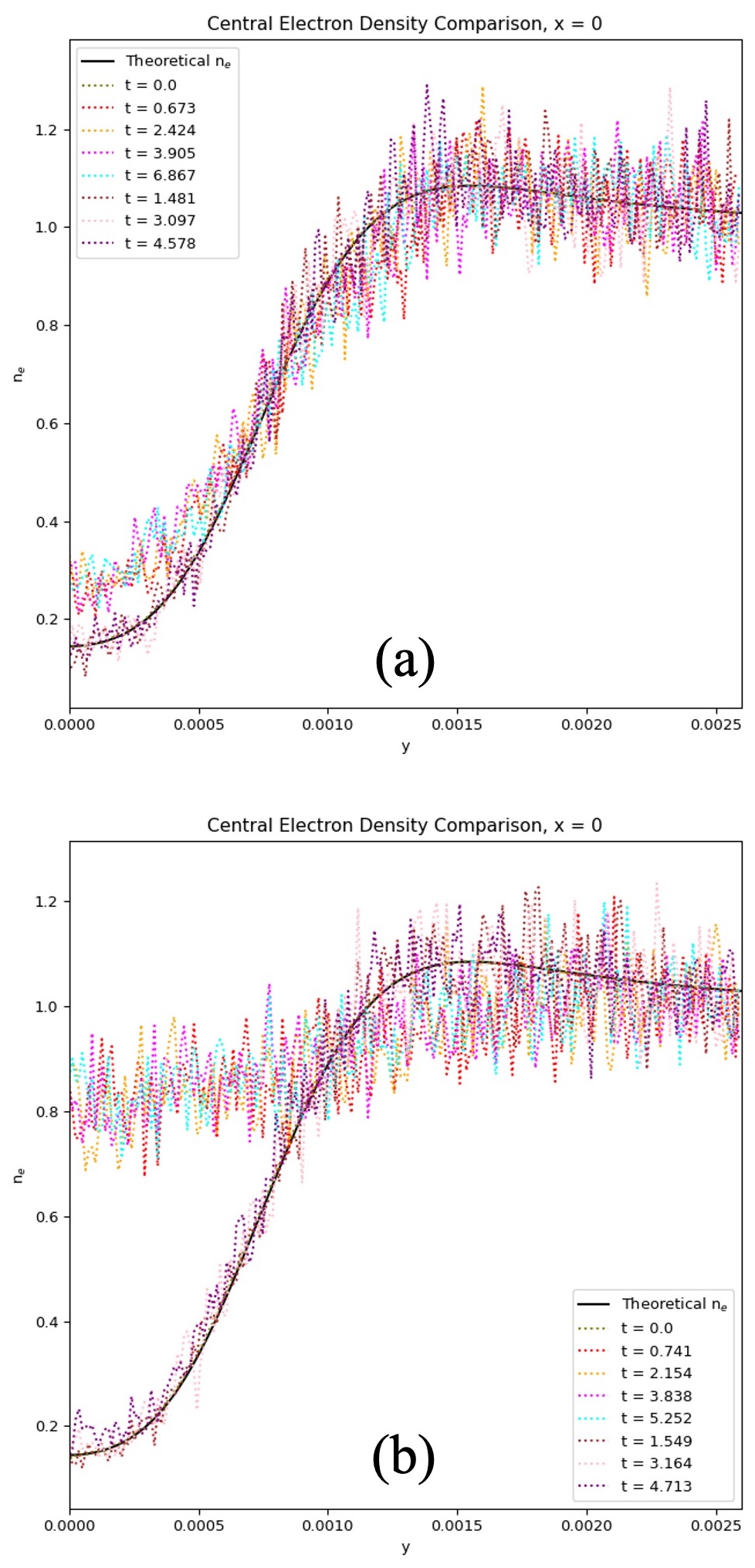}
    \caption{Electron density $n_e$ along the $y$-axis at indicated times from two PIC simulations with $B_0=4$ using a $2048^2$ grid, for (a) the normal-flow moment case, and (b) the reserved-flow moment case.
        The black continuous curves on both panels are
          the input $n_e$ profiles for the analytic BGK
          mode.
        \label{fig:moment-ne}}
\end{figure}

We now look at a specific case with $B_0=4$ to
  compare the moment runs with normal and reversed
  flow in more details.
Fig.~\ref{fig:moment-ne} shows curves of $n_e$ vs
  $y$ along the $y$-axis at different instants
  within about three $\tau_c = 2\pi/\omegac =
      \pi/2$, for the normal case in panel (a) and the
  reversed case in panel (b).
These instants are chosen when the $n_e$ value at
  the origin are roughly the largest or smallest
  during one $\tau_c$.
Therefore, $n_e$ curves between two instants are
  somewhere in between the corresponding two groups
  of curves that are shown.
The black continuous curves on both panels are
  the input $n_e$ profiles for the analytic BGK
  mode.

PIC simulations are known to suffer from noise
  due to finite particle number.
\cite{10.1063/1.862452}
The curves in Fig.~\ref{fig:moment-ne}
  demonstrate this, despite being from runs with
  our largest grid size of $2048^2$ cells.
The noise is low enough that the pulsation
  amplitude for both normal and reversed case is
  resolved.
While the noise level is about the same for the
  normal and reversed cases, the reversed case has
  a much larger pulsation amplitude.
Based on this result and similar results for
  every other relevant case, we conclude that the
  difference in behavior is based on true dynamics,
  not numerical noise.

The fact that the normal flow case has a small
  pulsation amplitude, as compared with the
  reversed flow case, as well as the initial
  depletion of $n_e$ at the center, shows that the
  BGK-like initial condition does represent a more
  time-steady and stable case.
By BGK-like, we mean this configuration only
  matches the three lowest-order moments (density,
  flow velocity, and effective temperature) of the
  true BGK mode solution.
Such a configuration cannot represent the
  non-Maxwellian distribution as shown in
  Fig.~\ref{fig:reduced dist}, especially for low
  $B_0$.
Nevertheless, the above results do show that even
  just matching the moments can make a
  configuration approximately time-steady in the
  strong $B_0$ limit.

We have put many simulation movies into the
  Supplemental Materials, including the two density
  movies corresponding to Fig.~\ref{fig:moment-ne}.
The movies for these two runs provide a better
  contrast between the different pulsation levels
  for the two cases.

\subsubsection{General Trends}

\begin{figure}
    \centering
    \includegraphics[width=\columnwidth]{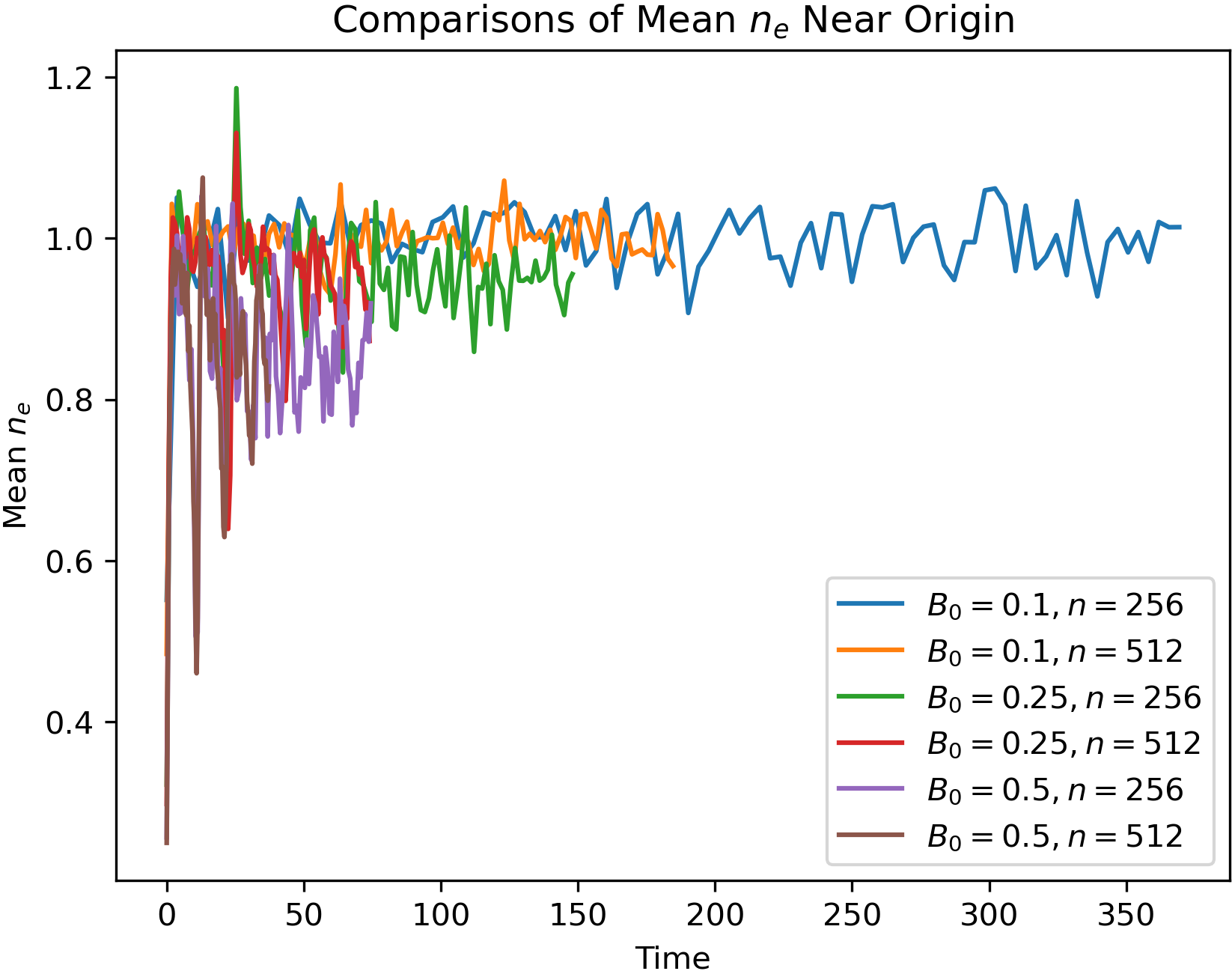}
    \caption{Electron density $n_e$ at the center for $B_0 = 0.1$, 0.25, and $0.5$ using $256^2$ and $512^2$ grids for moment runs.\label{fig:ne-center-weak}}
\end{figure}

\begin{figure}
    \centering
    \includegraphics[width=\columnwidth]{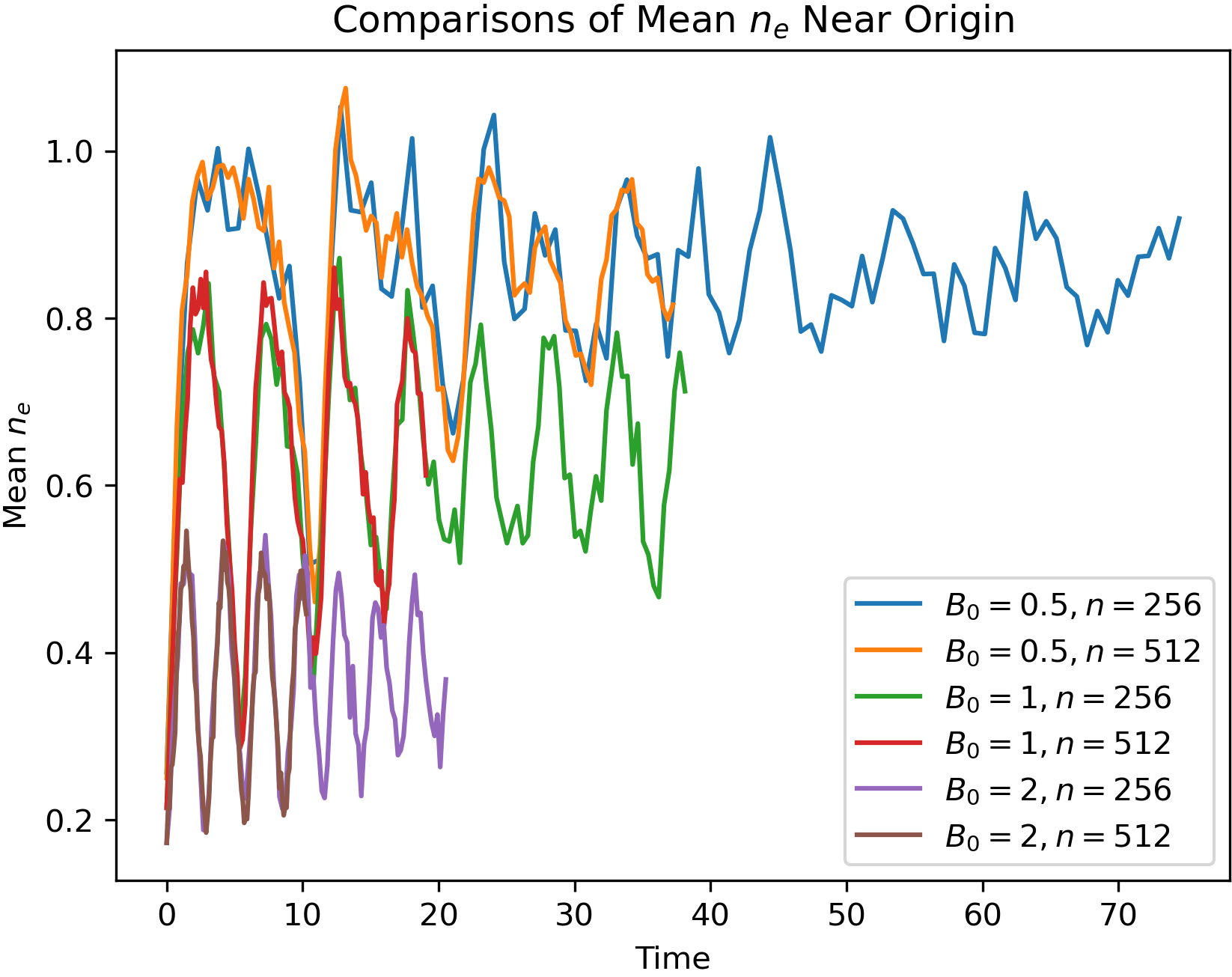}
    \caption{Electron density $n_e$ at the center for $B_0 = 0.5$, 1, and 2 using $256^2$ and $512^2$ grids for moment runs.\label{fig:ne-center-medium}}
\end{figure}

\begin{figure}
    \centering
    \includegraphics[width=\columnwidth]{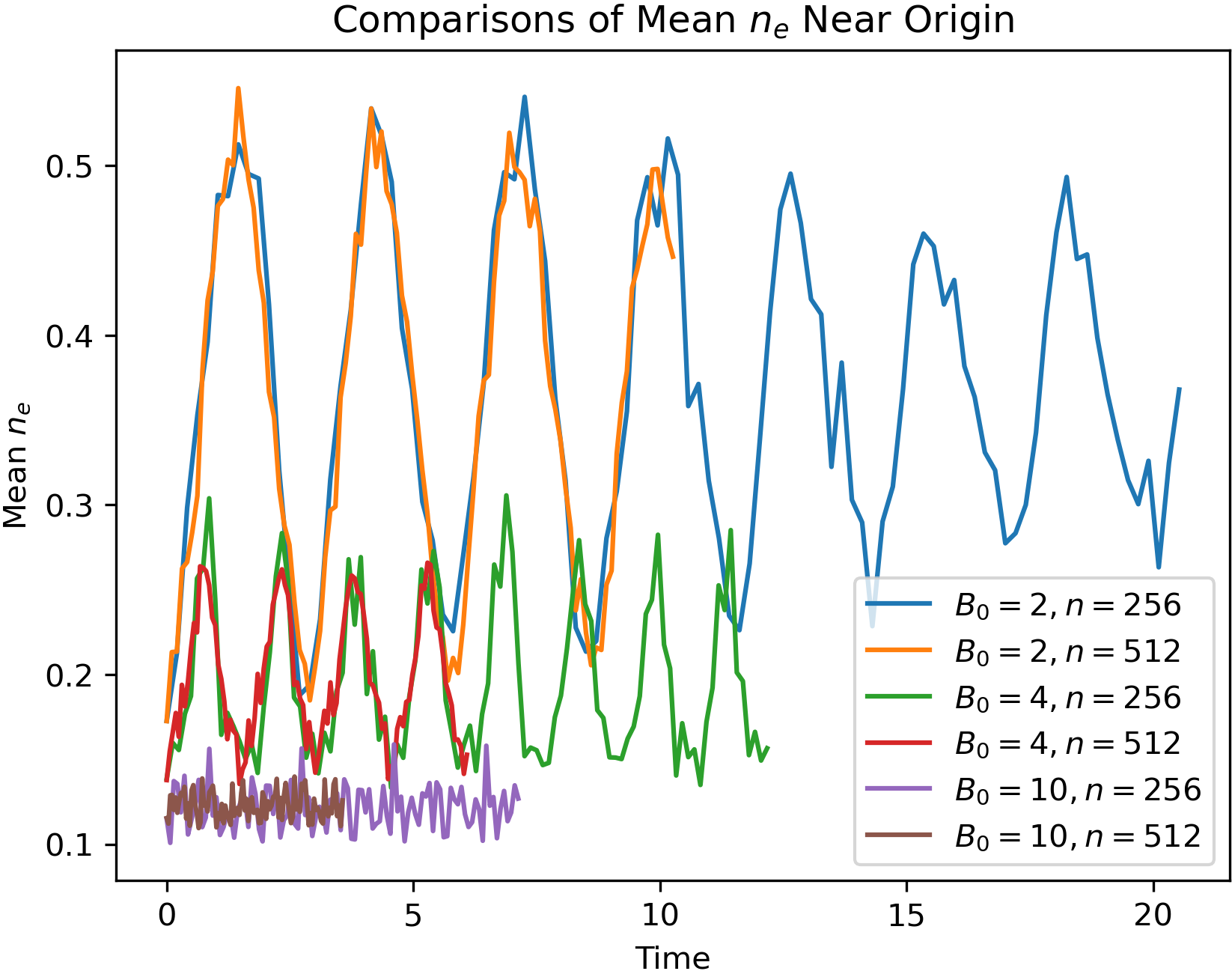}
    \caption{Electron density $n_e$ at the center for $B_0 = 2$, 4, and 10 using $256^2$ and $512^2$ grids for moment runs.\label{fig:ne-center-strong}}
\end{figure}

Generally, the pulsation level is larger for
  smaller $B_0$, and much larger for the reversed
  case than the normal case.
The $n_e$ value at the center can be greater than
  unity at times for smaller $B_0$ cases.
This value as a function of time turns out to be
  a simple visualization to show the pulsation of a
  run.
Figs.~\ref{fig:ne-center-weak},
  \ref{fig:ne-center-medium}, and
  \ref{fig:ne-center-strong} show the spatial
  averaged $n_e$ value at the center, $n_{e0}$, as
  functions of time, for $B_0$ in the weak, medium,
  and strong ranges (with overlaps).

In
  \Cref{fig:ne-center-weak,fig:ne-center-medium,fig:ne-center-strong},
  we show two sets of curves using $256^2$ and
  $512^2$ grids to show the possible dependence on
  resolution.
The figures indicate that the plasma behavior is
  essentially the same for the two grid
  resolutions.
Therefore, the dynamics of the moment runs are
  adequately resolved using a resolution of $256^2$
  or higher.

\subsubsection{Pulsation Frequency\label{sec:pulsation frequency}}

Angular frequencies of the pulsating behavior of
  moment runs are given in Table \ref{tab:freqs}.
These values were obtained by taking a Fourier
  transform of the mean electron density in the
  middle 4 cells of the $256^2$ runs, or an
  equivalent-sized square area for
  higher-resolution runs.
The angular frequency $\Omega$ is then $2\pi$
  times the frequency with the largest amplitude.
Although minor, pulsation for high-$B_0$ moment
  runs were detectable using this method.

\begin{table}[htb]
    \caption{\label{tab:freqs}
        Angular oscillation frequencies $\Omega$ of
          moment and moment reversed runs on $256^2$ grids,
          or $512^2$ grids for $B_0<1$.
    }
    \begin{tabular}{lrr}
        $B_0$ & $\Omega$ & $\Omega_\text{rev}$ \\
        \hline
        0.1   & ---      & ---                 \\
        0.25  & 0.320    & 0.232               \\
        0.5   & 0.635    & 0.509               \\
        1     & 1.24     & 1.06                \\
        2     & 2.27     & 2.14                \\
        4     & 4.21     & 4.16                \\
        10    & 10.2     & 10.2
    \end{tabular}
\end{table}

The angular frequencies are very close to the
  electron gyrofrequency $\omega_{ce}$, especially
  for reversed runs and for runs with higher $B_0$.
The discrepancy can in part be explained by the
  effects of the electron $E \times B$ drift in a
  cylindrically symmetric radial electric field,
  which increase the effective gyrofrequency for
  non-reversed cases more than the reversed cases,
  since the electric field magnitude oscillates
  (even with sign reversal) for the latter.
This hints that the pulsation is due to
  gyromotion.

\begin{figure}[tbhp!]
    \centering
    \includegraphics[width=\linewidth]{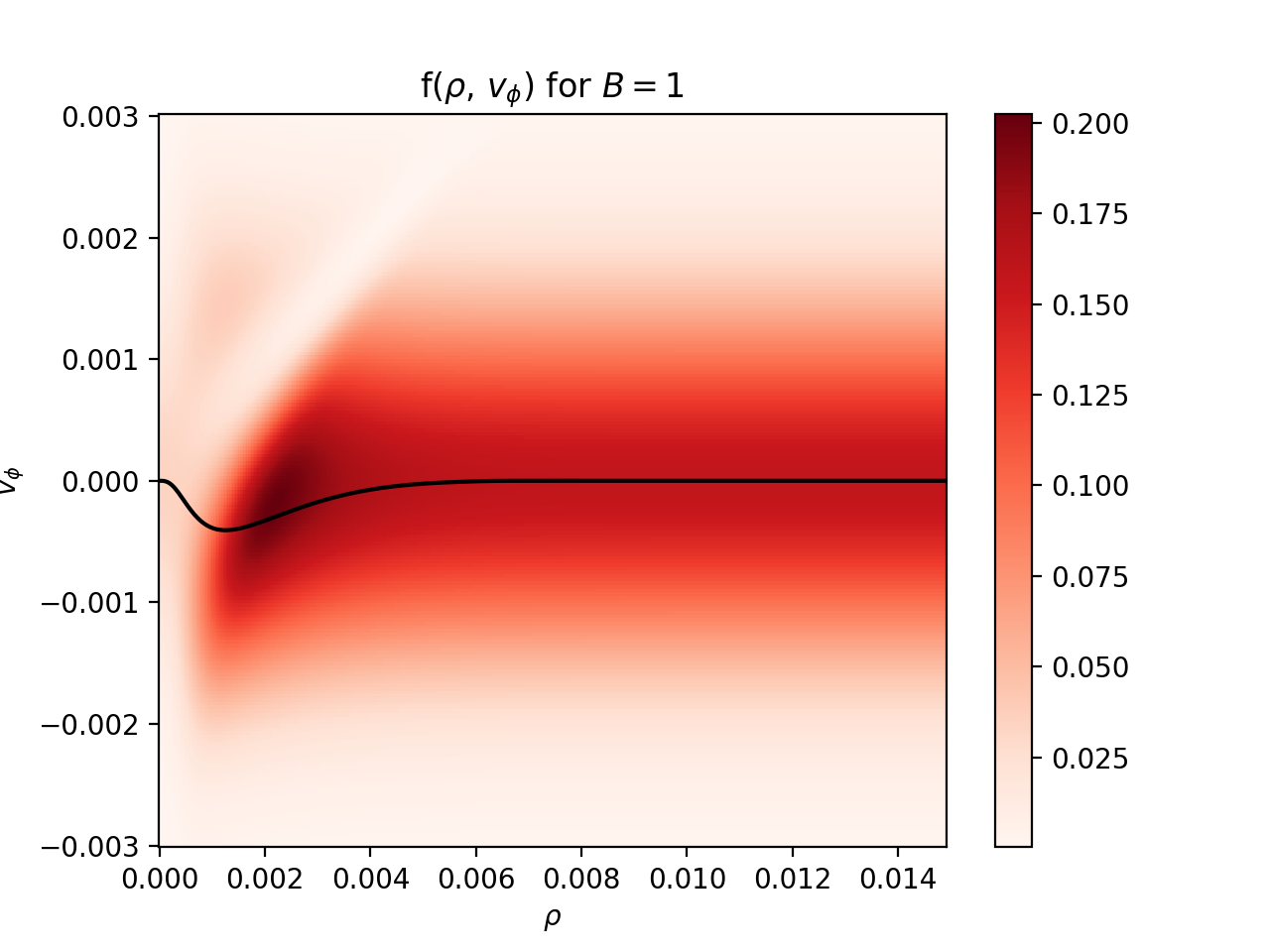}
    \caption{The reduced electron distribution $f(\rho, v_\phi)$ has a depleted ``spike" region extending through the $(\rho, v_\phi)$ space.
        The spike is steeper for higher $B_0$,
          disappearing into the hole itself.
        \label{fig:exact_dist}}
\end{figure}

Another look at the velocity-space distributions
  in Fig.~\ref{fig:seqs_max} suggests that
  electrons gyrate between the ``spike" region and
  the central hole.
The spike region is present in
  Eq.~\eqref{eq:exact dist}, as
  Fig.~\ref{fig:exact_dist} shows and of which
  Fig.~\ref{fig:reduced dist} is a vertical,
  renormalized cross section.
Indeed, the spike region seems to lie on the line
  where an electron's gyrodiameter
  $2\abs{v_\phi}/B_0$ equals its radial coordinate
  $\rho$ and gyromotion would take an electron at
  $\rho$ inward to near the origin, and an electron
  near the origin outward to this $\rho$ location.
This analysis led in part to support in PSC for
  initialization of the true electron distribution
  function, as described in \cref{sec:vel init}.

\subsubsection{Spike Region\label{sec:spike}}

Rewriting Eq.~\eqref{eq:exact dist} in terms of $\rho, v_\rho,$ and $v_\phi$ reveals it to be the difference of two Gaussians (note the normalization is by Eq.~\eqref{eq:normalized 1}, not Eq.~\eqref{eq:normalized 2}):
\begin{multline}
    \frac{\int \dd v_\rho\dd v_z f}{\int \dd^3 v f} = \frac1\alpha \frac{1}{\sqrt{2\pi}} e^{-\frac12v_\phi^2} \\
    + \left(1-\frac1\alpha\right)\frac{\gamma}{\sqrt{2\pi}} e^{-\frac12\gamma^2\left(v_\phi - {kB_0\rho^3}/{\gamma^2}\right)^2}
\end{multline}
where
\begin{align}
    \alpha \ideq \alpha(\rho) & \defeq 1-\frac h\gamma e^{-{kB_0^2\rho^4}/{4\gamma^2}}.
\end{align}

The spike is caused by the shifted Gaussian
  having a negative coefficient, i.e.,
  $1-1/\alpha<0$.
The negative Gaussian is centered at
  $v_\phi=kB_0\rho^3/\gamma^2$, which for $\rho>1$
  (or in PSC units: $\rho>.001$) gives $v_\phi\apeq
      \frac12 B_0\rho$.
Ignoring the effect of $v_\rho$, which is
  normally distributed with a mean of 0, this is
  the condition for an electron to have a
  gyrodiameter equal to its distance from the
  origin and to gyrate towards the origin.

The spike itself is not centered at the negative
  Gaussian, but slightly higher, due to the
  influence of the positive Gaussian.
This shift is such that not only does
  $v_\phi\to\frac12 B_0\rho$, but $\rho v_\phi \to
      \frac12 B_0\rho^2$, i.e., $l\to0$.
Note that the mean of the negative Gaussian does
  not satisfy the latter condition.

\subsection{Exact Runs}

\begin{figure*}
    \centering
    \begin{subfigure}[b]{\textwidth}
        \centering
        \includegraphics[width=\textwidth]{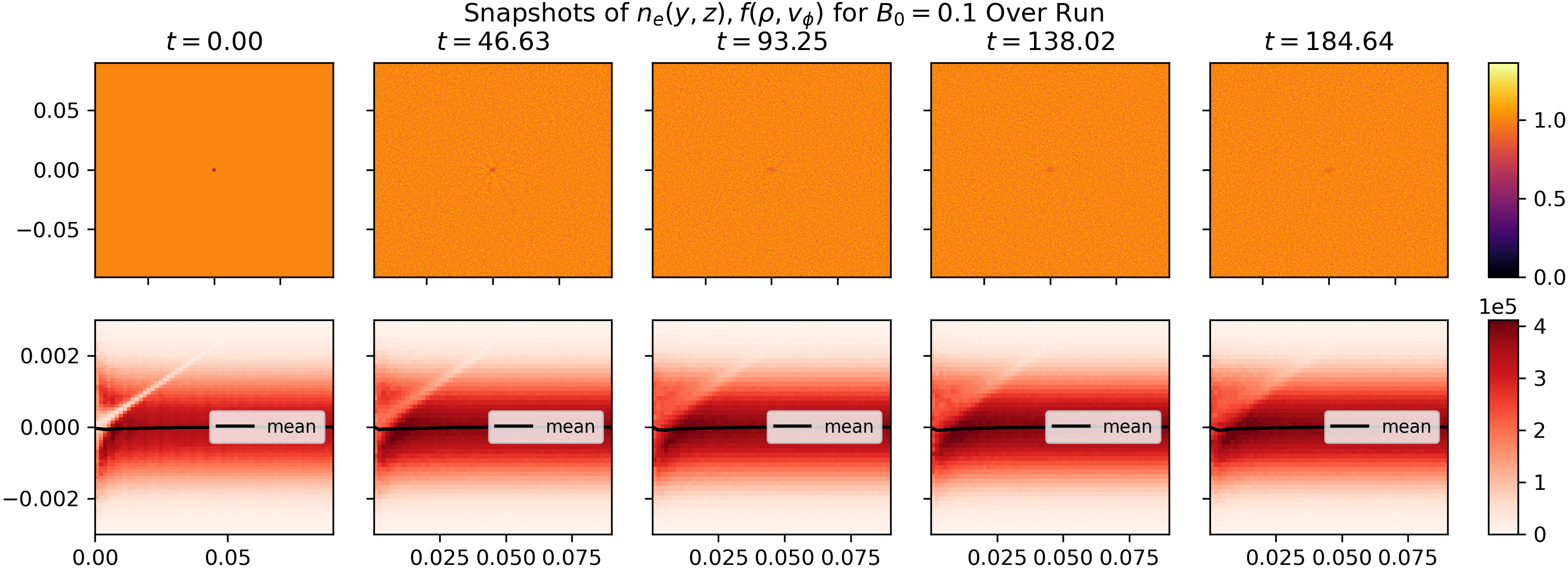}
    \end{subfigure}
    \begin{subfigure}[b]{\textwidth}
        \centering
        \includegraphics[width=\textwidth]{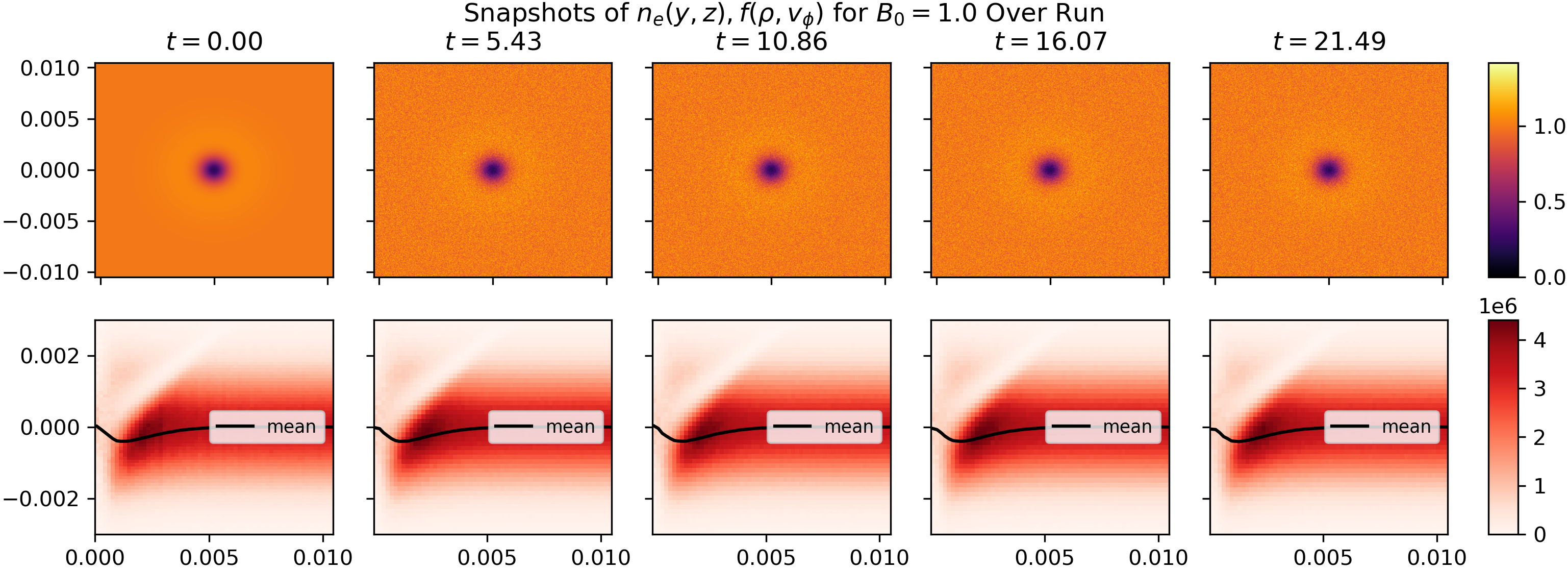}
    \end{subfigure}
    \begin{subfigure}[b]{\textwidth}
        \centering
        \includegraphics[width=\textwidth]{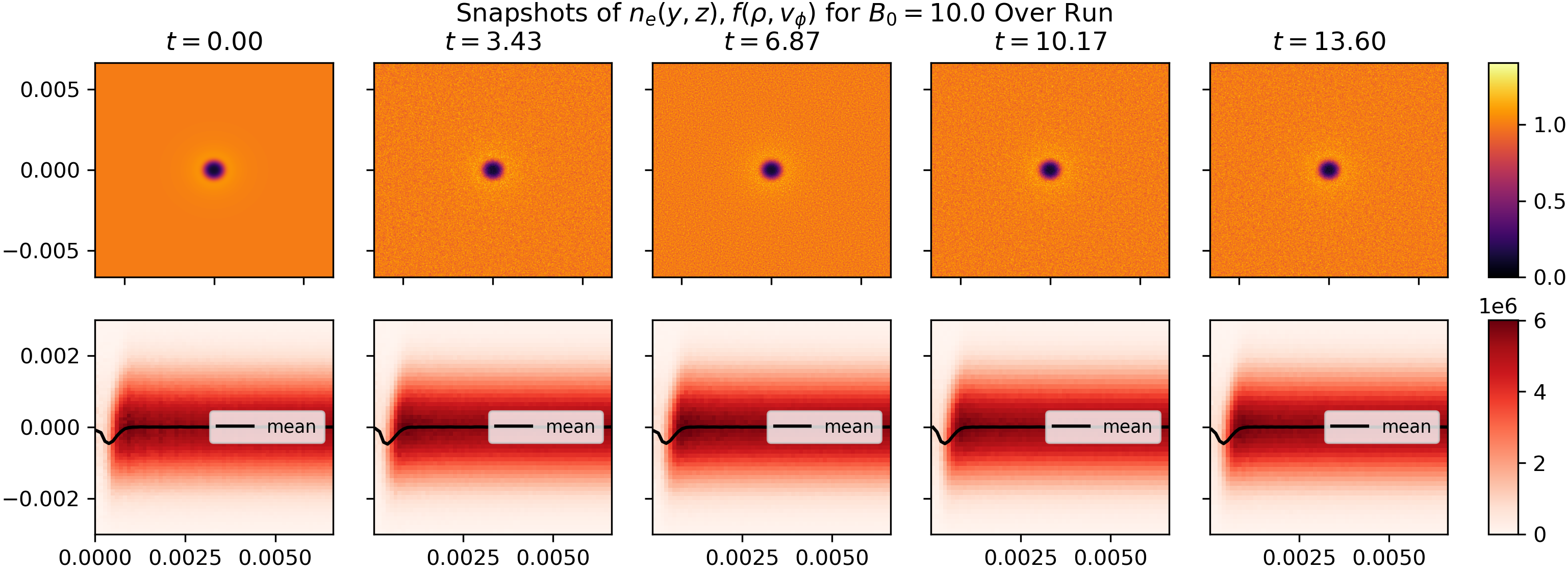}
    \end{subfigure}
    \caption{Sequences of $f(y,z)$ and $f(\rho, v_\phi)$ for ``exact" runs. \label{fig:seqs_exact}}
\end{figure*}

Exact runs, initialized according to the true
  electron distribution function, did not pulsate
  like the moment runs.
Figure \ref{fig:seqs_exact} gives snapshots of
  low-, mid-, and high-$B_0$ exact runs.
Note the spike regions are present at $t=0$, and
  both the density hole and velocity-space spike
  persists somewhat even for low-$B_0$ cases.

Periodicity was not detectable using the FFT
  method described in \cref{sec:pulsation
      frequency}.
Videos do seem to exhibit periodicity, however:
  macroparticles return to their initial
  configuration at the end of every gyroperiod,
  briefly returning the system to a relatively
  smooth state in between noisy periods.

Exact reversed runs behaved similarly to moment
  reversed runs and do not merit further
  discussion.

\subsubsection{Truly Stable: $B_0=4$, Exact}

With the pulsating behavior gone, we were able to
  determine whether the analytic form was
  time-steady or not, and if so, whether it was
  stable or unstable.
We use $B_0=4$ as an example again to contrast it
  with the results discussed in
  \cref{sec:stable-moment}, and because the
  magnetic field was strong enough that the
  configuration was stable.

\begin{figure*}
    \centering
    \includegraphics[width=\textwidth]{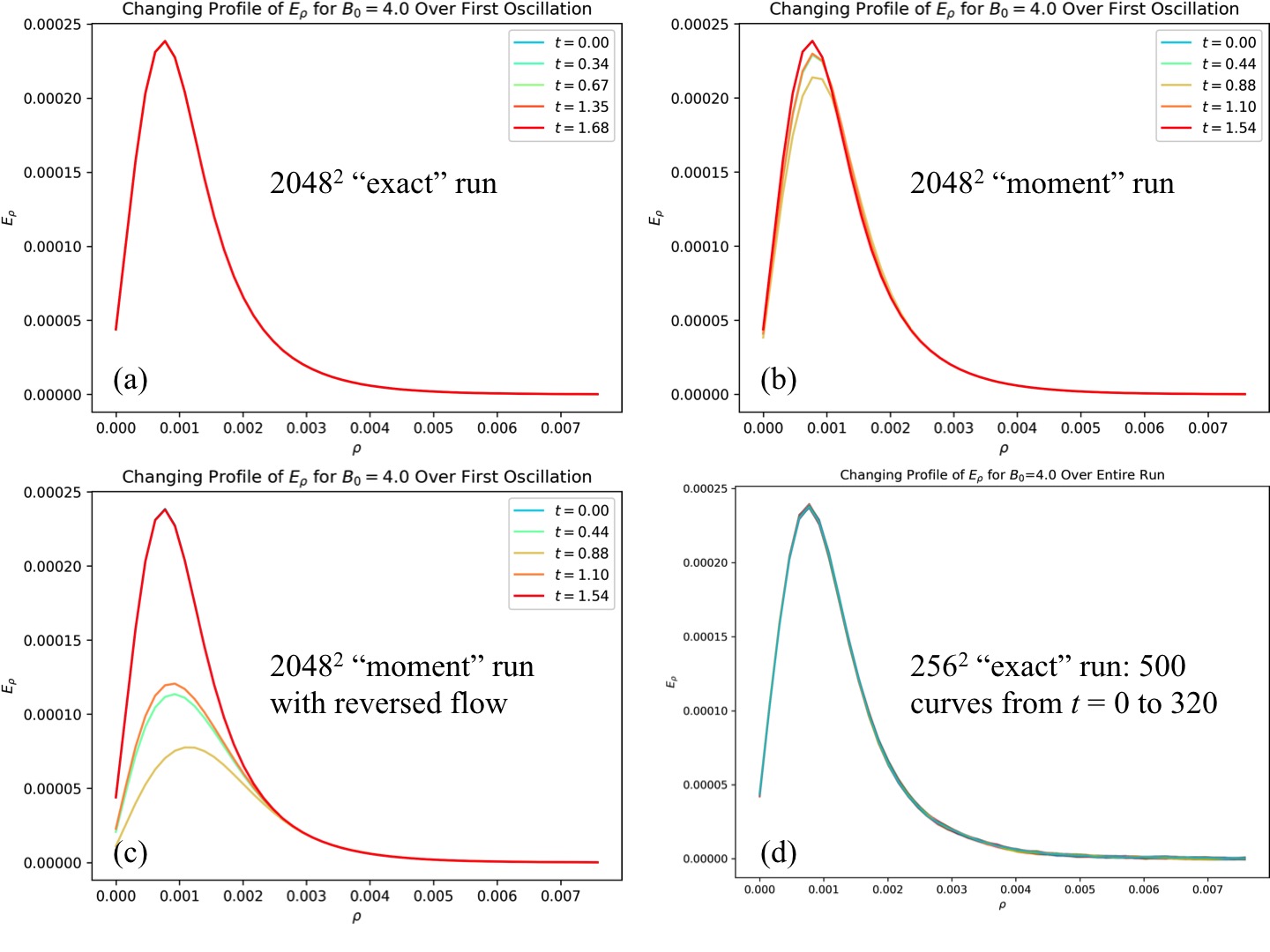}
    \caption{Profiles of the radial electric field component $E_\rho$, averaged over all azimuthal angles, as functions of $\rho$, for (a) an exact $2048^2$ run, (b) a moment $2048^2$ run, (c) a moment reversed $2048^2$ run, and (d) an exact $256^2$ run.
        Five curves each are plotted in the first three
          panels, over about one $\tau_c$.
        500 curves are plotted in (d), equally spaced over the entire duration of the run. \label{erho-plots}}
\end{figure*}

Figure~\ref{erho-plots} shows plots using data
  from four runs with $B_0=4$.
In addition to the moment and moment reversed
  runs using $2048^2$ grids from
  Fig.~\ref{fig:moment-ne}, we show a $2048^2$ run
  initialized exactly.
Its total duration is again not very long---up to
  $t \sim 6.3$, or about 4 $\tau_c$---since it is
  computationally expensive to run at such a high
  resolution.
To see the long term behavior, we performed
  another exact $B_0 = 4$ run using a $256^2$ grid
  and ran it up to $t \sim 320 \sim 204 \tau_c$.

To show the time-steadiness of the exact runs, we
  plot the radial electric field component
  $E_\rho$, averaged over all azimuthal angles, as
  a function of $\rho$, the radial position.
This averaging removes the random noise due to
  finite grid size (as seen in
  Fig.~\ref{fig:moment-ne}) but preserves the
  actual dynamics of time evolution.
The moment and reversed moment runs effectively
  serve as controls of this process.

From Fig.~\ref{erho-plots}(a) we see that the
  high-resolution exact run is indeed time-steady
  over one $\tau_c$, to the extent that all five
  $E_\rho$ profiles are on top of each other.
One the contrary, profiles of $E_\rho$ for the
  high-resolution moment and moment reversed runs,
  shown in Fig.~\ref{erho-plots}(b) and (c), have
  significant variations within one $\tau_c$,
  except that the profiles at the end of one
  $\tau_c$ come back to the initial profiles.
To demonstrate that the solution is actually
  stable, Fig.~\ref{erho-plots}(d) shows 500
  $E_\rho$ profiles for the $256^2$ exact run over
  204 $\tau_c$.
Except for a few small deviations, these 500
  curves are also almost on top of each other.
The time-steadiness of the two exact runs can be
  seen more clearly from the movies in the
  Supplementary Material.

\subsubsection{Other Stable Cases}

Our simulations with $B_0=2$ and $B_0=10$ were
  similar to those with $B_0=4$.
The $B_0=10$ case is extremely steady; even the
  corresponding moment run only slightly pulsated.
The $B_0 = 2$ cases also look stable generally,
  but the electron density hole is observed to have
  slight shifts or deformations in a long duration
  $256^2$ exact run.

\begin{figure}
    \centering
    \includegraphics[width=\columnwidth]{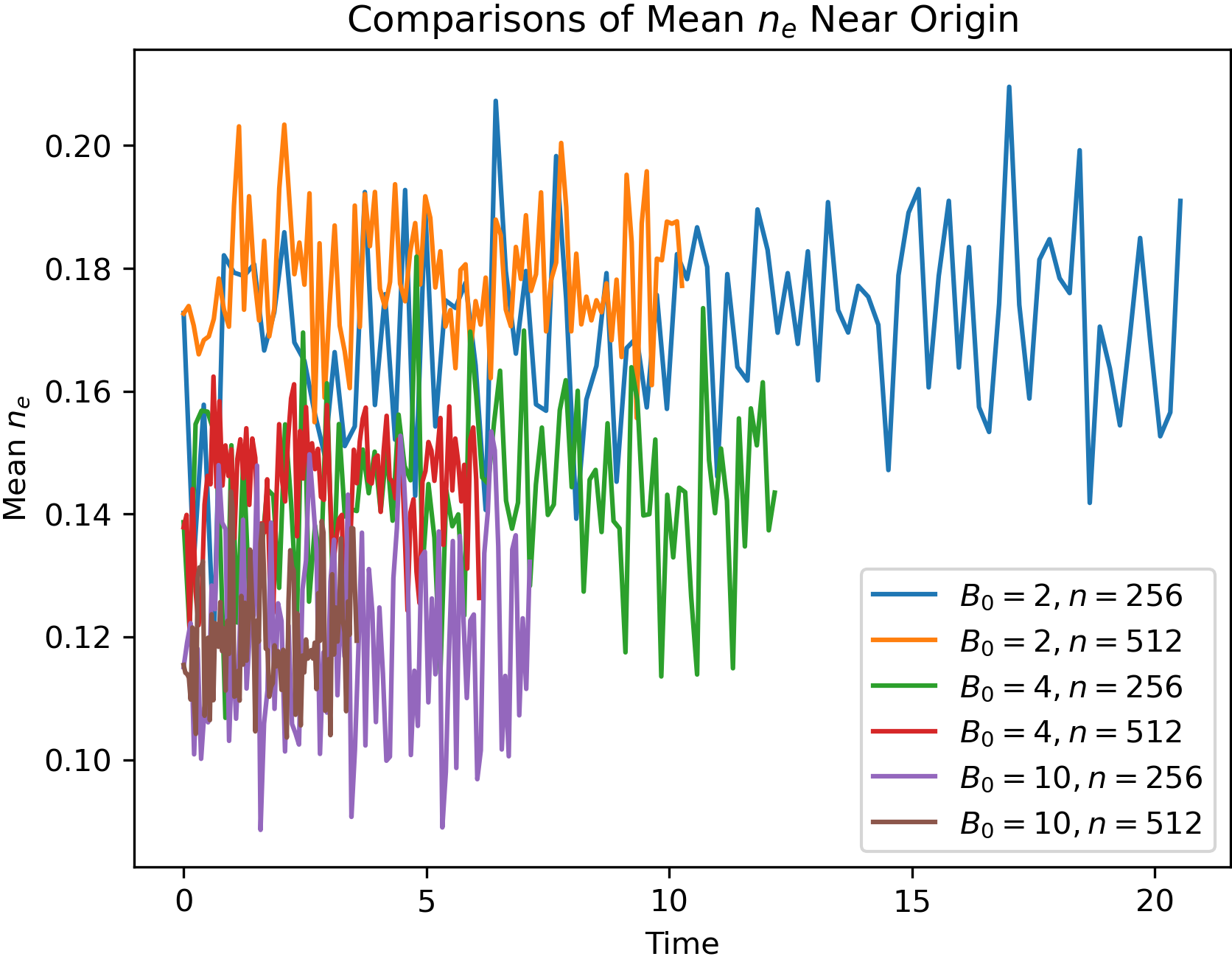}
    \caption{Electron density $n_e$ at the center for $B_0 = 2$, 4, and 10 using $256^2$ and $512^2$ grids for exact runs.\label{ne0-exact-strong}}
\end{figure}

Figure~\ref{ne0-exact-strong} shows the $n_{e0}$
  values for exact runs using these three values of
  $B_0$.
Compared to the corresponding curves of
  Fig.~\ref{fig:ne-center-strong}, $n_{e0}$
  fluctuates with smaller amplitudes in exact runs
  than moment runs for $B_0 = 2$ and 4 cases.
The fluctuations are ostensibly comparable for
  the $B_0 = 10$ cases, but the $B_0 = 10$ moment
  run has clear pulsation in the $n_e$ movie,
  whereas the movie for the exact run does not.

To the best of our knowledge, this is the first
  time 2D BGK mode solutions of a form similar to
  Eq.~(\ref{eq:exact dist}) have been confirmed
  with a kinetic simulation.
This is remarkable in the sense that such
  solutions are fully nonlinear, self-consistent
  localized solutions of the Vlasov-Poisson system
  of equations, i.e., Eqs.~(\ref{eq:vlasov}) and
  (\ref{eq:poisson}).
Moreover, this result also shows that there
  exists solutions that are stable over some ranges
  of parameters, at least with perturbations
  confined in the 2D plane, and up to rather long
  durations of the simulations.
The stability study for more general
  perturbations using 3D PIC simulations is
  ongoing, and will be the subject of a later
  publication.

\subsubsection{Unstable Cases}

\begin{figure}[tbhp]
    \centering
    \includegraphics[width=\linewidth]{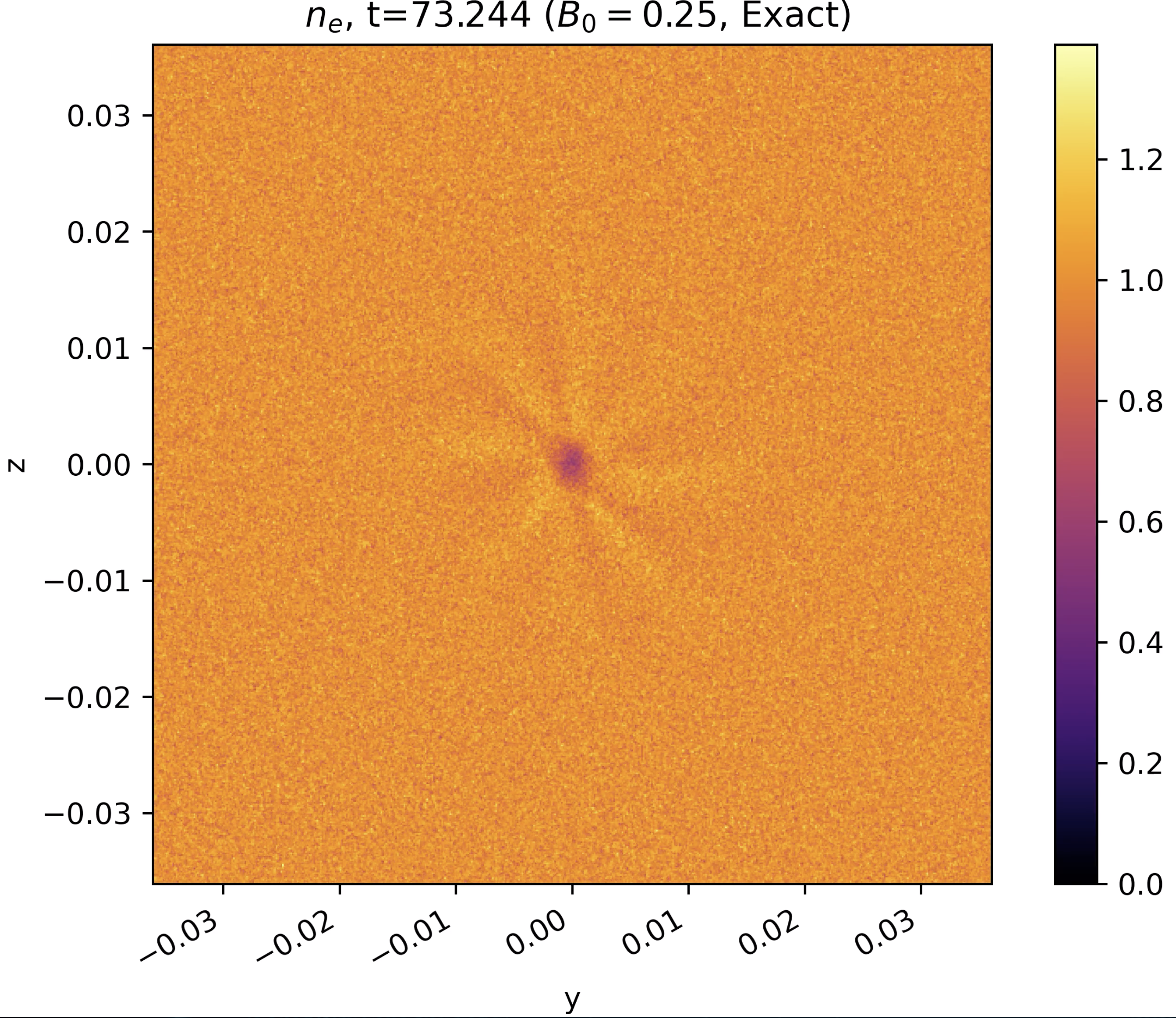}
    \caption{$n_e$ map for an exact $B_0=0.25$ run of resolution $512^2$ at the given time.\label{fig:spiral-ne}}
\end{figure}

Exact runs with $B_0\lesssim 1$ were unstable.
Unstable modes can appear due to the bimodal
  velocity distribution as in Fig.~\ref{fig:reduced
      dist}.
The instability manifests as radial arms in $n_e$
  that rotate counterclockwise around the origin,
  as shown in Figs.~\ref{fig:spiral-ne}; note that
  the electron flow is clockwise.

The fluctuations are especially visible in the
  azimuthal component of the electric field
  $E_\phi$, as in \ref{fig:spiral view} and in the
  movies in the Supplementary Material.
We thus identify these rotating fluctuations as
  electrostatic waves.
The propagation appears as roughly co-rotating at
  first, and then either forms a spiral pattern as
  it fades away.

Spiral waves are clearly seen in the $n_e$ and
  $E_\phi$ movies for $B_0 = 0.25$, 0.5, and 1.
For the $B_0 = 0.1$ case, the initial wave has
  many more arms than other cases (12 or more), but
  the co-rotation phase quickly ends with the
  almost total dissipation of the wave, obscuring
  any possible spiral phase and making analysis
  difficult.

For the $B_0 = 1$ case, the instability takes a
  long time ($t > 100$) to appear, but the spiral
  wave lasts a long time, persisting until we had
  to stop the simulation at $t=433$.
This case, as well as the $B_0 = 0.5$ case, has
  only four arms, while the $B_0 = 0.25$ case has
  between 6 and 8.
Some of these arms are not present over the full
  range of $\rho$ for which there are unstable
  waves; rather, there appears to be bifurcations
  such that one arm can split into two when going
  radially outwards, as can be seen in
  Fig.~\ref{fig:spiral view}.
Generally, not all arms are of the same strength,
  suggesting a mixture of unstable modes instead of
  one single mode.

Given the spiral structure of the density
  fluctuations, the waves should also have $E_\rho$
  fluctuations such that the wave has a component
  propagating outwards along the $\rho$ direction.
The $E_\rho$ fluctuations are more difficult to
  see because $E_\rho$ from the background BGK mode
  is large compared with the fluctuations.
We include one $E_\rho$ movie in the Supplemental
  Materials for the $B_0 = 1$ case to illustrate
  this point.
In fact, the wave starts to have more apparent
  outward propagation when the spiral pattern
  emerges, and this could be the physical reason
  behind the formation of the spiral pattern.

\begin{figure}[tbhp]
    \centering
    \includegraphics[width=\linewidth]{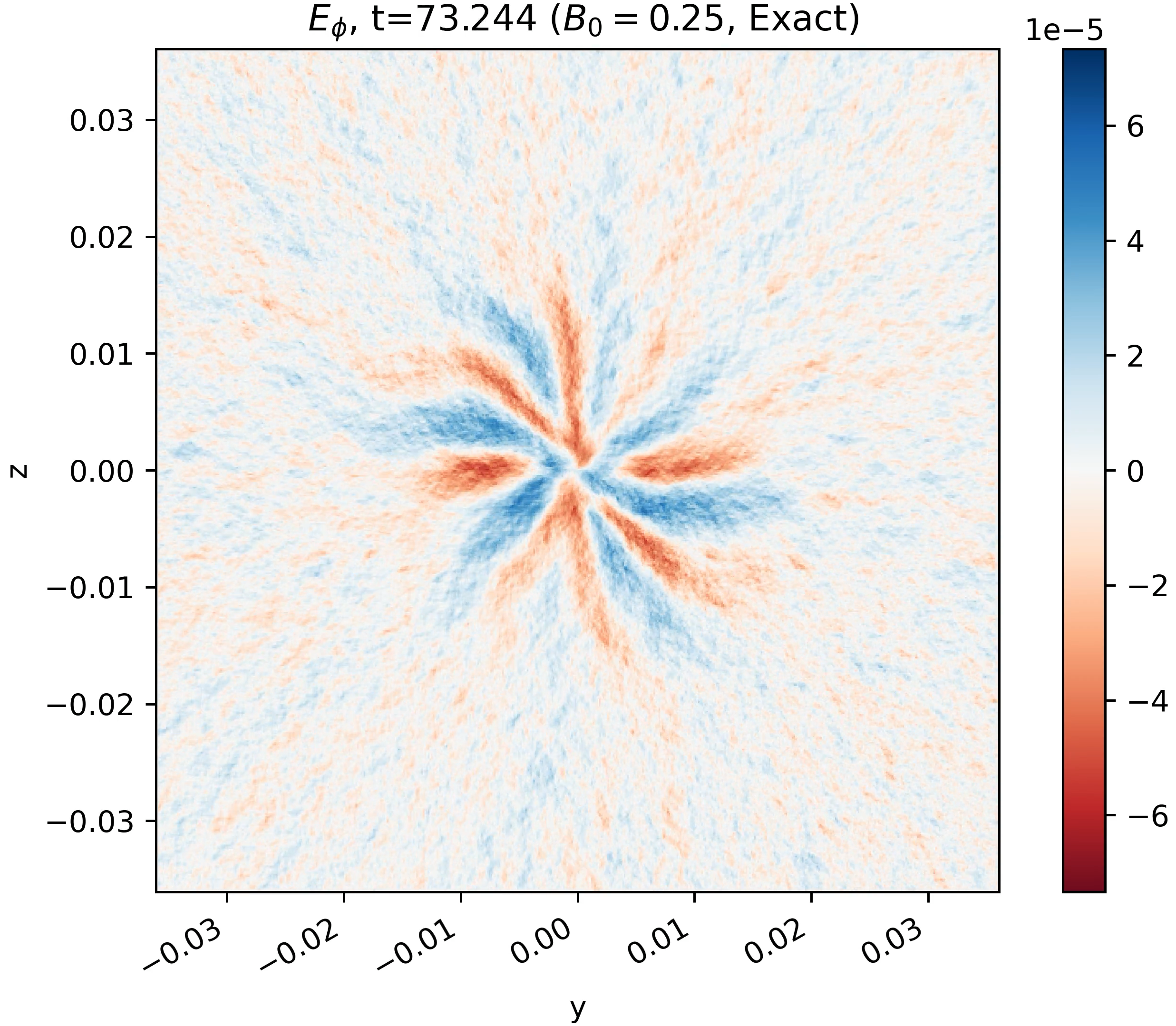}
    \caption{$E_\phi$ map for an exact $B_0=0.25$ run of resolution $512^2$at the given time.\label{fig:spiral view}}
\end{figure}

\subsubsection{Post-Instability}

\begin{figure}
    \centering
    \includegraphics[width=\columnwidth]{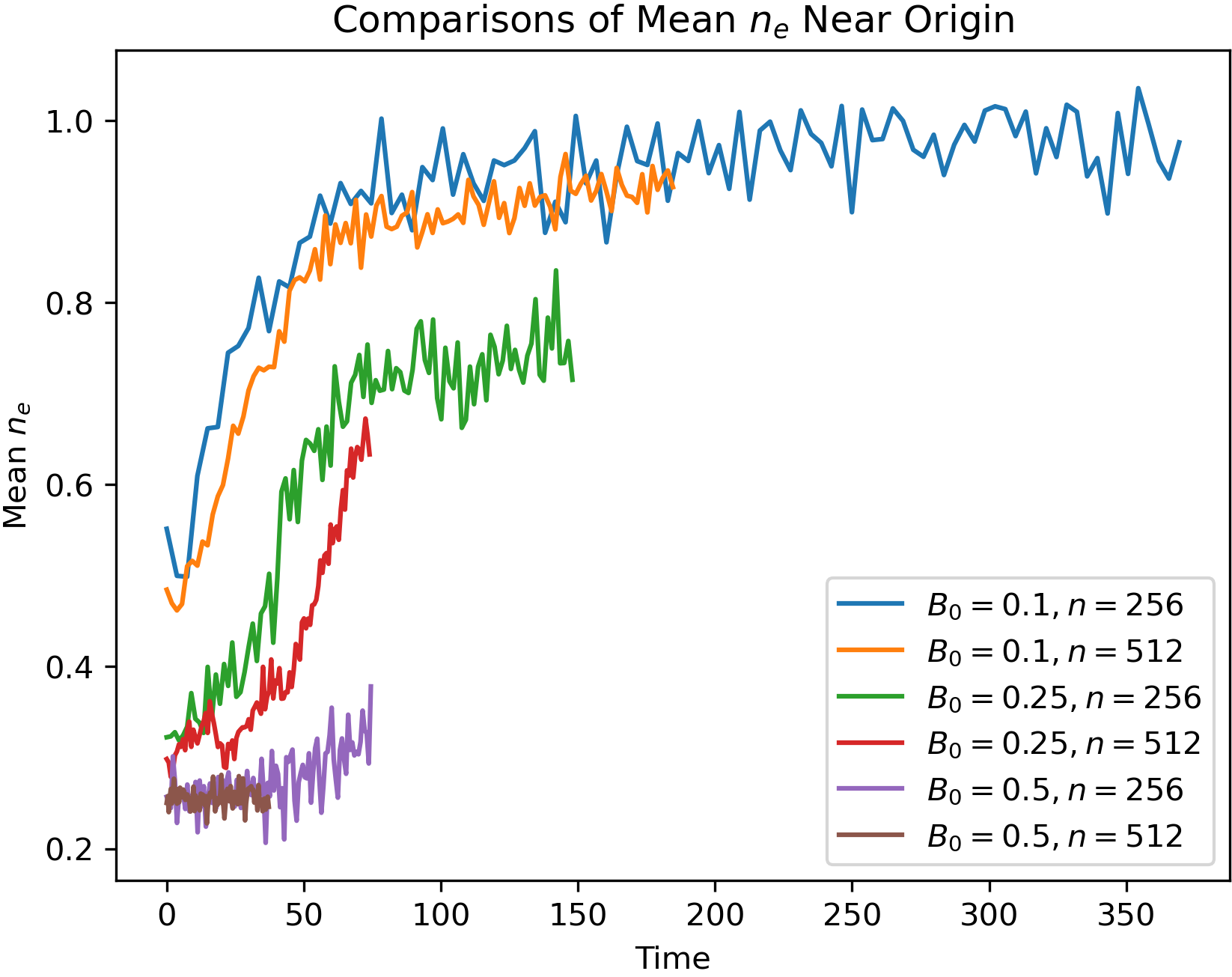}
    \caption{Electron density $n_e$ at the center for $B_0 = 0.1$, 0.25, and $0.5$ using $256^2$ and $512^2$ grids for exact runs.\label{ne0-exact-weak}}
\end{figure}

\begin{figure}
    \centering
    \includegraphics[width=\columnwidth]{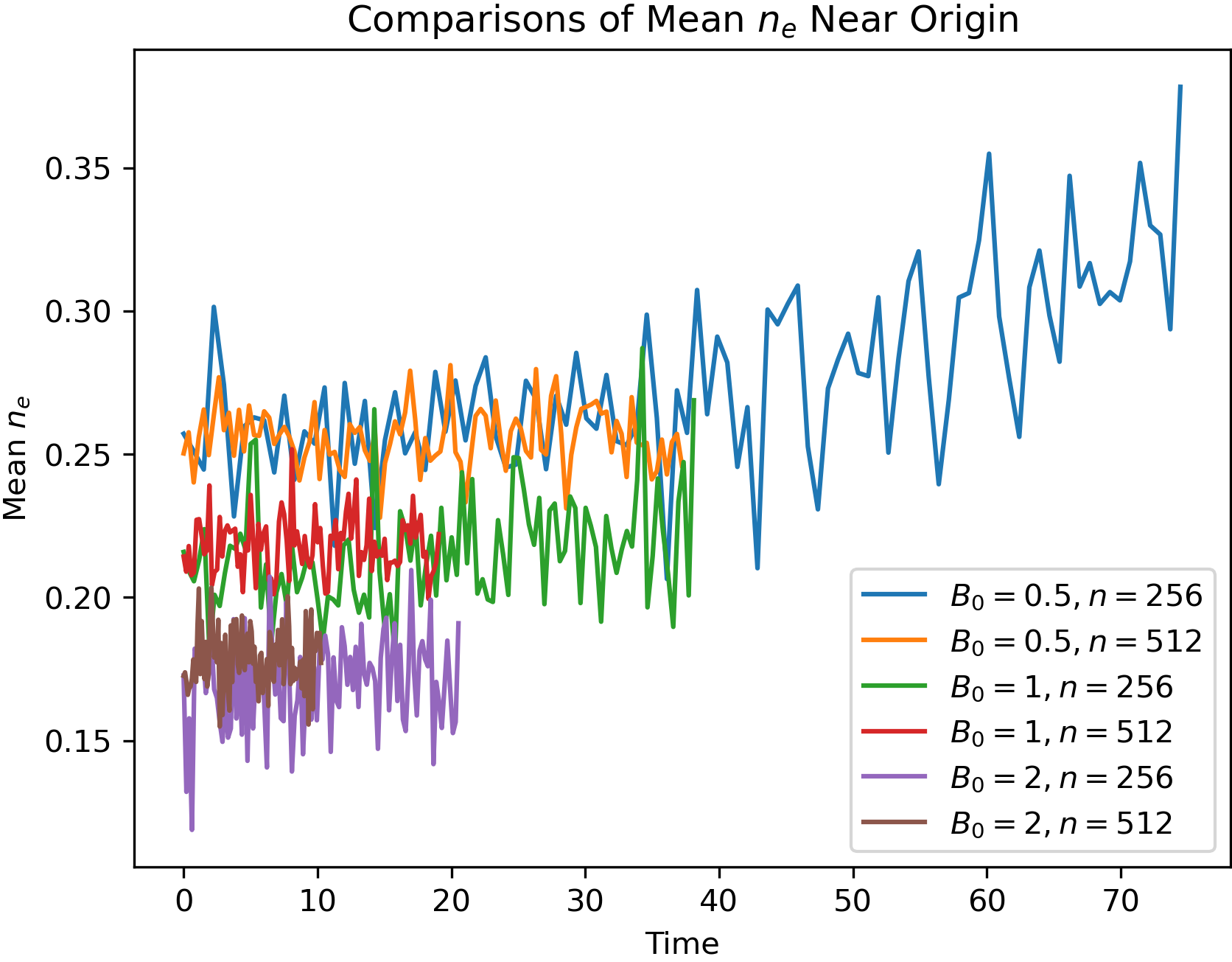}
    \caption{Electron density $n_e$ at the center for $B_0 = 0.5$, 1, and 2 using $256^2$ and $512^2$ grids for exact runs.\label{ne0-exact-medium}}
\end{figure}

It is observed that the electron density hole
  begins to fill as soon as the instability
  appears.
This can be seen from the $n_{e0}$ vs.
$t$ plots, e.g.
Fig.~\ref{ne0-exact-weak} for cases with weak
  $B_0$ and Fig.~\ref{ne0-exact-medium} for cases
  with medium $B_0$.

These two figures are significantly different
  from corresponding figures from the moment runs,
  i.e., Figs.~\ref{fig:ne-center-weak} and
  \ref{fig:ne-center-medium}.
The latter exhibit pulsation, while the former
  appear to follow logistic curves and saturate at
  a value below 1.
This is clearly show in Fig.~\ref{fig:logistic
      growths}.
We see that the growth of $n_{e0}$ starts near
  $t \sim 0$ for the $B_0 = 0.1$ and $0.25$ cases, while
  there may be a linear phase in the growth of
  $n_{e0}$ for the $B_0 = 0.5$ and 1 cases, as it
  takes some time for the instability to grow.

As far as our simulations can show, $n_{e0}$
  apparently stabilizes at a value below 1 for
  these unstable cases.
In other words, there is still an electron
  density hole after the initial hole decays,
  albeit one with an $n_{e0}$ closer to 1.
The spike also seems to survive the instability.
The snapshots of the exact $B_0=1$ run in
  Fig.~\ref{fig:seqs_exact} show such a state in
  the last two frames, while the second frame
  captures a glimpse of the radial arms.
If an electron density hole truly does exist
  after the instability, it is unclear whether it
  would be in a stable time-steady state similar to
  the solutions described by Eq.~(\ref{eq:exact
      dist}).

\begin{figure}
    \centering
    \includegraphics[width=\columnwidth]{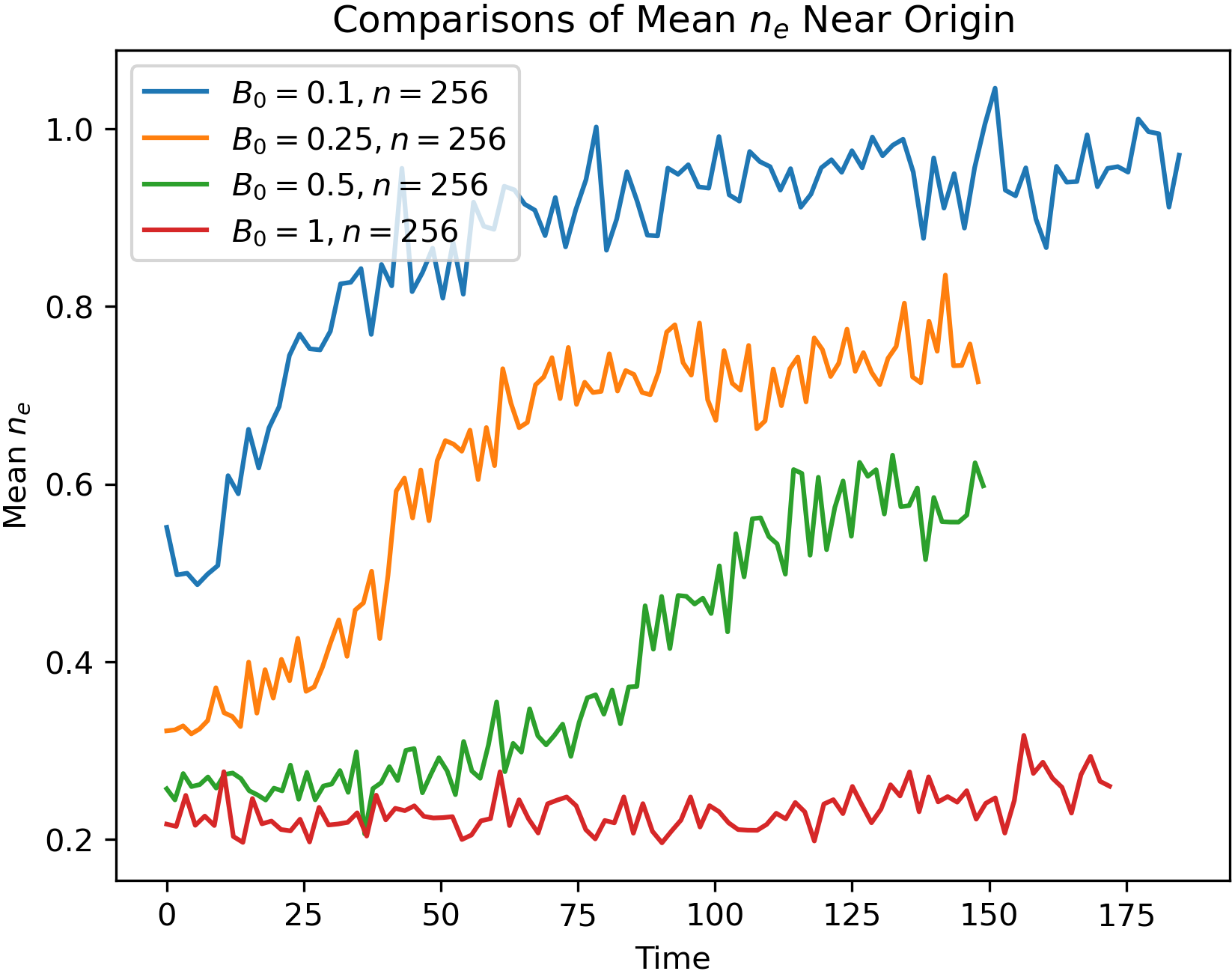}
    \caption{Electron density $n_{e0}$ at the origin for $B_0 \le 1$ using $256^2$ grids for exact runs.
        Higher-$B_0$ runs had to be extended multiple
          times to achieve such high times.
        \label{fig:logistic growths}}
\end{figure}

\subsubsection{Comments on Stability\label{sec:comments on stability}}

It is unknown what critical value of $B_0$
  separates stability from instability, since we
  have not simulated enough cases with different
  $B_0$.
The apparent stability of high-$B_0$ runs
  indicates that such a value exists, and our
  limited number of cases suggest that it is around
  $B_0 \sim 2$ for our choice of other parameters
  $h$ and $k$.
The chosen form of the analytic BGK mode
  distribution shown in Eq.~(\ref{eq:exact dist})
  implies that the critical value of $B_0$ depends
  on these parameters.
For example, using a smaller value of $h$ for the
  same $B_0$ will make the distribution more
  Maxwellian, and a Maxwellian distribution is well
  known to be stable.

Due to limitations of space and scope, we omit a
  more thorough analysis of the observed
  instabilities, including numerical results and a
  comparison to a theoretical model.
We will present such discussion in a separate
  publication instead.

\subsection{Scaling\label{subsec:scaling}}

\begin{figure}[tbhp]
    \centering
    \includegraphics[width=\linewidth]{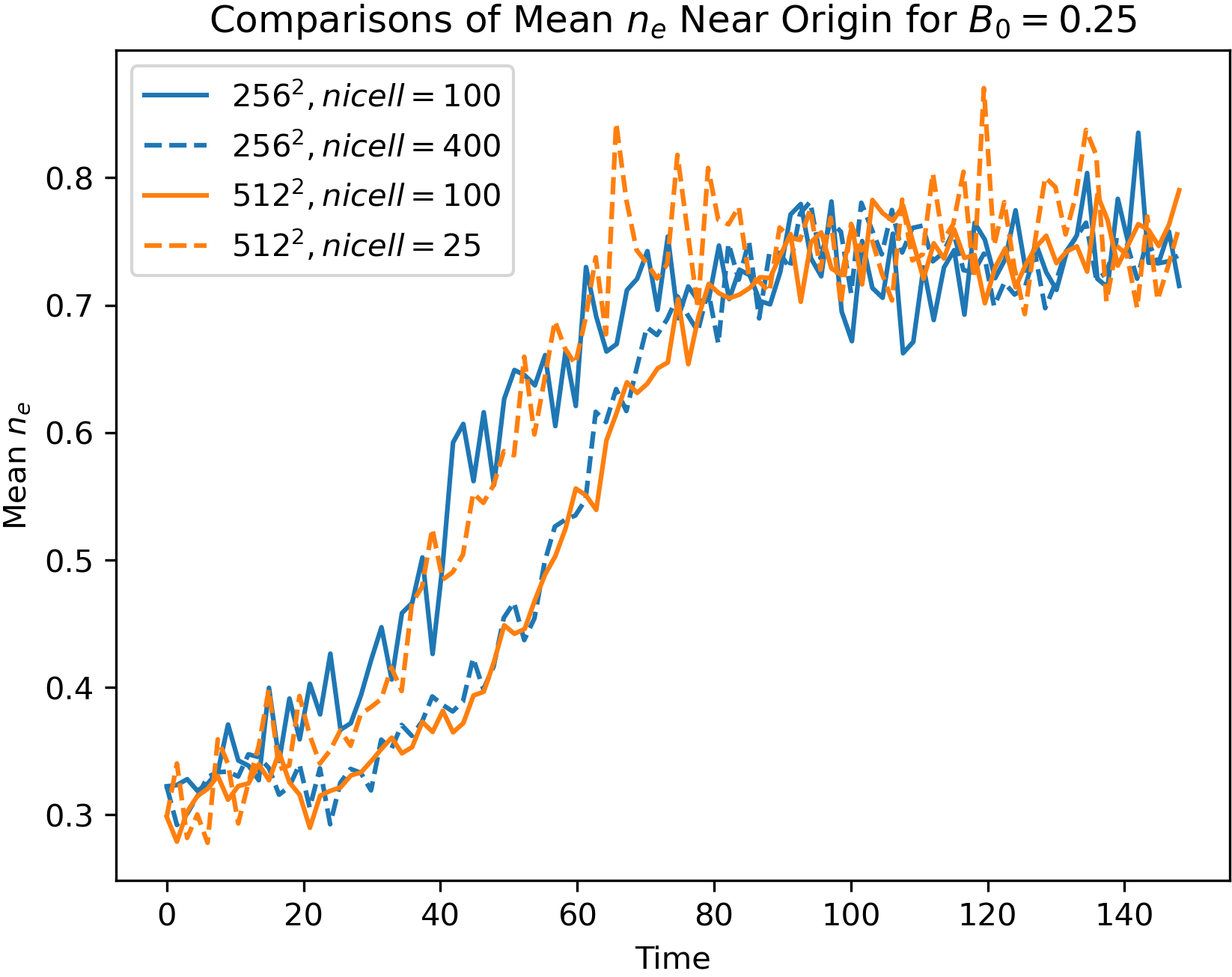}
    \caption{Comparing the effects of resolution and \code{nicell} on the growth rate of exact $B_0=0.25$ runs.
        Total number of electron macroparticles is
          $\code{nicell}*\text{number of cells}$.
        \label{fig:scaling vs nicell}}
\end{figure}

Unstable runs, i.e., exact runs with low $B_0$,
  were somewhat affected by resolution.
This was to be expected, since PIC simulations
  are inherently noisy, and it is likely that
  particle noise affects the evolution of the
  unstable configurations.
Figure~\ref{fig:scaling vs nicell} lends credit
  to this idea by showing that the instability
  growth depends more on \code{nicell} than on the
  resolution.
Importantly, although the total number of
  macroparticles affected the time for growth to
  begin, it significantly affected neither the
  growth rate nor the final $n_{e0}$ value.
It seems that particle noise kickstarts the
  growth process, and more noise triggers it
  sooner.

For the most part, other runs did not appear
  sensitive to scaling.
Noise was likely suppressed in high-$B_0$ exact
  runs, which were stable.
As for moment and reversed runs, the deliberate
  perturbations would have dominated any noise.
Plots similar to Fig.~\ref{fig:scaling vs nicell}
  were produced for these runs and showed
  convergence for all resolutions tested.

One possible exception is for cases with
  $B_0=0.1$.
For such cases, the spike extends far from the
  origin while the hole itself stays small.
Consequently, a large physical domain is needed
  to resolve the spike, so a high resolution is
  needed to resolve the hole.
In fact, the central 4 cells used to calculate
  $n_{e0}$ almost cover the entire density hole of
  the $256^2$ run, so the method used to plot the
  time series $n_{e0}$ is not very accurate.

Overall, while we did not have time to perform an
  extensive scaling analysis, we are reasonably
  certain that even our $256^2$ runs were
  sufficiently resolved---again, with the possible
  exception of $B_0=1$.
Certainly, our $2048^2$ runs were sufficiently
  resolved, and they exhibited the same qualitative
  behavior as our other runs.

\afterpage{\clearpage}
\section{CONCLUSION AND DISCUSSION \label{sec:conclusion}}

With the development of a new method to input an
  analytic distribution as initial conditions (the
  ``exact method'') in PSC, our simulations affirm
  the validity of Eq.~\eqref{eq:exact dist} as a
  steady-state solution to the Poisson-Vlasov
  system of equations.
This is the first time to our knowledge that such
  a solution has been validated using a kinetic
  code.
The evidence of time-steadiness was further
  supported by contrasting exact runs with
  ``moment'' runs, which effectively served as a
  control.
Moment runs were never as time-steady as their
  exact counterparts, even at high $B_0$ with
  resolutions of $2048^2$.
Since solutions of the form of
  Eq.~\eqref{eq:exact dist} are self-consistent,
  they may be useful as validation cases involving
  nonlinear physics for other kinetic codes, now
  that they are shown to be correct.

The validity of the time-steadiness of such
  solutions also shows that there exist 2D BGK
  modes that are stable for some ranges of
  parameters, under the restriction of 2D
  perturbations.
We must therefore take more seriously the
  possibility that these structures exist in
  nature.
To show this more conclusively, we will have to
  wait for more kinetic simulations in 3D, which we
  are working on currently.

Our PIC simulations also found that for our
  choice of parameters $h$ and $k$, solutions are
  stable in the large-$B_0$ limit, and unstable in
  the opposite limit.
In particular, the instabilities were found to
  excite azimuthally-propagating electrostatic
  waves.
These waves initially co-rotate, and can evolve
  into a spiral pattern as they start to decay.

The instability was also found to have the
  associated effect of filling the central electron
  density hole, although only partially: in every
  unstable case we ran, the electron density at the
  origin seemed to stabilize at a value less than
  1.
The resulting density hole may persist for a much
  longer period of time than the initial hole.

Moreover, except for very low values of $B_0$,
  the electron density holes in the moment (and
  even reversed moment) runs are only temporarily
  filled by pulsations.
We have not followed the long-time evolutions of
  such cases, since doing so would require more
  computation time than we could spare.
It would be of great interest to find out whether
  a generic electron density hole configuration
  would evolve to a new, stable hole configuration
  after instabilities and pulsations run their
  course, as we found with our weak-$B_0$ exact
  runs.
If the answer to this question is affirmative,
  this would provide a general mechanism for 2D BGK
  modes to form, further strengthening the
  possibility that they may exist in nature.

\subsection{Galactic Similarities}

The excitation of spiral waves from the
  instability of the 2D BGK modes could be of
  interest outside of plasma physics---galactic
  dynamics, specifically.
This is not because of the outward similarity of
  the spiral wave patterns to spiral galaxies, but
  because a collisionless plasma system and a
  stellar system are both described by the same
  equations: the Vlasov equation and the Poisson
  equation, with some sign changes to reflect the
  fact that the gravitational force between masses
  is always attractive, unlike electric forces.
In fact, like BGK modes, galaxies can be regarded
  as kinetic equilibria of this system of
  equations.
\cite{10.2307/j.ctvc778ff}
The fact that the physics behind the formation of
  spiral galaxies is considered somewhat
  unsolved\cite{10.2307/j.ctvc778ff} suggests that
  our observation of spiral waves excitation for
  unstable 2D BGK modes might provide some insight
  for that problem.

One might counter this suggestion by pointing out
  that our 2D solutions are actually tube-like when
  viewed in the 3D spatial domain, while a galaxy
  is a genuinely 3D object.
However, one of us (CSN) has developed a
  ``galactic disk'' model of a 3D BGK mode, to be
  presented in a separate publication, in which the
  form of the distribution of the disk particle is
  very similar to that of the 2D BGK modes, i.e.,
  Eq.~(\ref{eq:exact dist}).
It will be shown that a galactic disk can in fact
  have very similar physics to a 2D system.

\subsection{Future Work}

The research presented in this paper is our first
  attempt at studying the stability of 2D BGK modes
  using PIC simulations.
Therefore, we have started with the simplest
  cases, and have restricted the choices of
  parameters of the model, i.e., we kept
  $(k,h)=(0.4,0.9)$ fixed.
As discussed in \cref{sec:comments on stability},
  other choices of these parameters could change
  the stability boundary of the only parameter that
  we vary, i.e., $B_0$.

Additionally, we limited $v_e=c/1000$ to make
  sure that the theory can be meaningfully compared
  with simulations.
In our future studies, we plan to relax such
  restrictions progressively.
This would obviously require significant
  computational effort, and also more calculations
  as discussed in \Cref{sec:approximations}.
For example, a larger $v_e$ which is still
  non-relativistic would involve solving
  Amp\`{e}re's Law self-consistently alongside
  Eqs.~\eqref{eq:vlasov} and \eqref{eq:poisson}
  resulting in a BGK mode with nonuniform
  background magnetic field.
\cite{doi:10.1063/1.5126705}

One must also keep in mind that the form of the
  distribution given in Eq.~(\ref{eq:exact dist})
  is just one of infinitely many possible choices.
Thus, 2D BGK modes are really given by an
  unlimited number of parameters.
While this is an intimidating prospect, we can list a few interesting and meaningful generalizations that we can study in the future:
\begin{enumerate}
    \item choose a negative $h$ for an electric density ``bump'' instead of a hole at the center;
    \item choose a very small $k$ such that the mode will have a non-uniform magnetic field, generated self-consistently by Amp\`{e}re's Law even with a small but finite $v_e/c$;
    \item include the dependence on the axial component of the canonical momentum in the form of the distribution, such that the mode has an associate magnetic field that has a azimuthal component;\cite{doi:10.1063/1.5126705}
    \item include a background of ions that satisfies a Boltzmann distribution, and might not have the same temperature as the background electrons;
    \item consider BGK modes comprising kinetic ions with ion distributions depending on canonical angular momentum, with Boltzmann or kinetic electrons;
    \item and consider distributions with forms different from that of Eq.~(\ref{eq:exact dist}), such as with a linear dependence on $l$, instead of $l^2$, in the argument of the exponential function in the second term of the right-hand side.
\end{enumerate}

Although this list is by no means exhaustive, it
  would clearly require a huge effort to address
  all the tasks properly, and cannot possibly be
  finished by our team alone.
This discussion is simply to demonstrate the rich
  possibilities of different kinds of small kinetic
  structures, and it is likely that new and
  unexpected physics can come from some of them.


%
%

%

\begin{acknowledgments}
    This work is supported by National Science
      Foundation grants PHY-2010617 and PHY- 2010393.
    This research used resources of the National
      Energy Research Scientific Computing Center, a
      DOE Office of Science User Facility supported by
      the Office of Science of the U.S.
    Department of Energy under Contract No.
    DE-AC02-05CH11231.
\end{acknowledgments}

\section{Supplementary Material}

https://drive.google.com/drive/folders/12DLFK-kR7upP-wHh0V4Pjyc1Y3veVIah


\providecommand{\noopsort}[1]{}\providecommand{\singleletter}[1]{#1}%

\end{document}

%% file: preamble.tex
\usepackage{amsmath,amssymb,amsthm}
\usepackage{enumitem}
\usepackage{mathtools}
\usepackage{esvect}
\usepackage[margin=1in]{geometry}
\usepackage{ifthen}

\newcommand{\defeq}{\mathrel{\coloneqq}}

\let\ideq\equiv
\let\apeq\approx

\DeclarePairedDelimiter{\abs}{\lvert}{\rvert}

\newcommand{\dd}{\text{d}}
\newcommand{\deriv}[2]{\frac{\dd #1}{\dd #2}}
\newcommand{\pderiv}[2]{\frac{\partial #1}{\partial #2}}

\renewcommand{\vec}[1]{\protect\vv{#1}}
\let\unit\hat

\let\code\texttt

\newtheorem*{axiom*}{Axiom}

\newtheorem*{defn*}{Definition}

\newtheorem*{theorem*}{Theorem}

\newtheorem*{metatheorem*}{Metatheorem}

\newtheorem*{lemma*}{Lemma}

\newtheorem*{cor*}{Corollary}

\newtheorem*{remark*}{Remark}

\newtheorem*{prop*}{Proposition}